\documentclass[a4paper,11pt]{article}
\usepackage[toc,page]{appendix}
\usepackage{graphicx}
\usepackage{colortbl}
\usepackage{amsmath}
\usepackage{subfigure}
\usepackage{mathtools}
\usepackage[utf8]{inputenc}
\usepackage{float}
\usepackage[normalem]{ulem}
\usepackage{epstopdf}
\usepackage{amsfonts,amsmath,amssymb}
\usepackage{hyperref}

\usepackage{epsfig,multicol,bbm}
\usepackage{color}
\usepackage[dvipsnames]{xcolor}

\definecolor{darkblue}{rgb}{0.0, 0.0, 0.55}
\definecolor{grey}{rgb}{0.57, 0.64, 0.69}
\definecolor{lightbrown}{rgb}{0.71, 0.4, 0.11}

\newcommand{\be}{\begin{equation}}
\newcommand{\ee}{\end{equation}}

\newcommand\fverb{\setbox\pippobox=\hbox\bgroup\verb}
\newcommand\fverbit{\egroup\item[\fbox{\unhbox\pippobox}]}

\newbox\pippobox

\begin{document}

 \title{Analytically Approximation Solution to Einstein-Cubic Gravity}
 \author{S. N. Sajadi\thanks{Electronic address: naseh.sajadi@gmail.com}\,,\, S. H. Hendi\thanks{Electronic address: hendi@shirazu.ac.ir}
\\
\small Department of Physics, School of Science, Shiraz University, Shiraz 71454, Iran \\
\small Biruni Observatory, School of Science, Shiraz University, Shiraz 71454, Iran\\
\small Canadian Quantum Research Center 204-3002 32 Ave Vernon, BC V1T 2L7 Canada }

\maketitle
\begin{abstract}
In this work, we introduce analytical approximate black hole
solutions in Einstein-Cubic gravity. To obtain complete solutions,
we construct the the near horizon and asymptotic solutions as the
first step. Then, the approximate analytic solutions are obtained
through continued-fraction expansion. We also compute the
thermodynamic quantities and use the first law and Smarr formula
to obtain the analytic solutions for near horizon quantities.
Finally, we follow the same approach to obtain the new static
black hole solutions with different metric functions.
\end{abstract}

\section{Introduction}

Higher-order gravity models recently attracted considerable
attention. In the context of cosmology, in order to go beyond the
standard $\Lambda$CDM model and find an explanation for the
late-time accelerated expansion, dark matter or inflation
\cite{Sotiriou:2008rp}-\cite{Nojiri:2010wj}, higher-order curvature
gravity theories are helpful. In AdS/CFT context,
higher-order gravities have been used as tools to characterize
numerous properties of strongly coupled conformal field theories
\cite{Maldacena:1997re}-\cite{Camanho:2009hu}. From quantum gravity
viewpoint, in order to unify quantum mechanics and gravitational
interactions, going beyond the Einstein gravity is necessary
\cite{Stelle:1976gc}.

In recent years, a new class of higher derivative theories has been discovered that is ghost-free and in four dimensions neither topological nor trivial known as Generalized Quasi-Topological Gravity \cite{Myers:2010ru}-\cite{Ahmed:2017jod}.
One of the such higher-derivative gravity theories which in the four
dimensions is neither topological nor trivial is Einsteinian cubic
gravity. This theory of gravity, that has been recently proposed
in \cite{Bueno:2016xff}, is the most general up to cubic order in
curvature dimension independent theory of gravity that shares its
graviton spectrum with Einstein's theory on constant curvature
backgrounds. The Einsteinian cubic gravity field equations admit
generalizations of the Schwarzschild solution, i.e. static,
spherically symmetric solutions with a single metric function
\cite{Bueno:2016lrh,Hennigar:2018hza,Hennigar:2016gkm}. The
Lagrangian density of this theory is given by
\cite{Bueno:2016lrh,Poshteh:2018wqy}
\begin{equation*}
L=R-2\Lambda +\beta_{1}\kappa_{4}+\beta_{2}\kappa_{6}+\alpha P,
\end{equation*}
where $\kappa_{4}$ and $\kappa_{6}$ are four and six-dimensional
Euler densities and correspond to the usual Lovelock terms, and
$P$ is the cubic term. In 4-dimensions the terms proportional to
$\beta_{1}$ and $\beta_{2}$ have no contribution on the field
equations. In \cite{Bueno:2016lrh}-\cite{Frassino:2020zuv}, the authors construct static and
spherically symmetric generalizations of the Schwarzschild and
Reissner-Nordstr\"{o}m-(Anti-)de Sitter black-hole solutions in
four-dimensions and study the orbit of massive test bodies near a
black hole, especially computing the innermost stable circular
orbit. They compute constraints on the ECG coupling parameter and
the shadow of an ECG black hole. In \cite{Feng:2017tev}, bounce
universe in the critical point of the coupling constants of the
theory has been studied. In \cite{Jiang:2019kks}, the holographic
complexity of AdS black hole in Einsteinian cubic gravity has been
investigated through the "complexity equals action" and
"complexity equals volume" conjectures. In \cite{Edelstein:2022xlb} the condensation of a charged scalar field in a (3 + 1)-dimensional asymptotically AdS background in the context of Einsteinian cubic gravity has been studied. In \cite{Bueno:2018yzo} holography on squashed-spheres, in \cite{Bueno:2018xqc} the holographic entanglement entropy, and in \cite{Bueno:2018xqc} various aspects of holographic ECG have been studied.
In
\cite{Poshteh:2018wqy}, the gravitational lensing due to the
presence of supermassive black holes at the center of the Milky
Way and other galaxies in Einsteinian Cubic Gravity has been
studied.

In this paper, by using the continued fraction expansion
technique, we obtain the static spherically symmetric solutions
for the theory. This ansatz is designed so that the coefficients in the continued fraction are fixed by the behavior of the metric near the event horizon, while the pre-factors are introduced to match the asymptotic behavior at infinity \cite{Kokkotas:2017zwt},\cite{Konoplya:2019ppy},\cite{Zinhailo:2018ska}. This method
is an accurate analytic method that has recently been applied with
success in a variety of contexts
\cite{Rezzolla:2014mua},\cite{Sajadi:2020axg}. These analytic
studies of the black hole also allowed us to study thermodynamics
and other properties of the solutions. While, the numerical solutions do not give a clear picture of the metric dependence on physical parameters of the system.

The paper is organized as follows: In the next section, we first
review Einsteinian cubic gravity and continued-fraction expansion.
Then, calculating the thermodynamical quantities and inserting
them in the first law and Smarr formula, we obtain the solutions
for the near horizon quantities. We also plotted the metric functions and thermodynamical  quantities and compared them with the previous works on this theory,  we find a good agreement between them. In Sec. \ref{sec3}, we introduce new solutions of
the theory with different metric functions. Finally, we conclude
the paper in Sec. \eqref{con}.

\section{Basic equations}\label{sec2}

In 4D, Einstein cubic gravity theory is determined by the action \cite{Bueno:2016lrh}-\cite{Frassino:2020zuv}
\begin{equation}\label{eq1}
\mathcal{S}=\dfrac{1}{16\pi G}\int d^{4}x\sqrt{-g}(R-2\Lambda +\alpha P-\kappa F_{a b}F^{a b}),
\end{equation}
where $R$ represent the Ricci scalar, $\alpha$ is coupling constant of the theory, and $P$ cubic-in-curvature correction to the Einstein-Hilbert action is given as \cite{Bueno:2016lrh}-\cite{Frassino:2020zuv}
\begin{equation}
P=12R_{a}{}^{c}{}_{b}{}^{d}R_{c}{}^{e}{}_{d}{}^{f}R_{e}{}^{a}{}_{f}{}^{b}+R_{a b}{}^{c d}R_{c d}{}^{e f}R_{e f}{}^{a b}-12R_{abcd}R^{ac}R^{bd}+8R_{a}{}^{b}R_{b}{}^{c}R_{c}{}^{a}.
\end{equation}
The correction term in four dimensions is dynamical and is not topological or trivial \cite{Bueno:2016lrh}-\cite{Frassino:2020zuv}.
Using the variational principle, one can find the following
equation of motion
\begin{align}\label{fieldeq3}
E_{a b}&=P_{a c d e}R_{b}{}^{c d e}-\dfrac{1}{2}g_{ab}L-2\nabla^{c}\nabla^{d}P_{a c d b}-2T_{a b}=0,\nonumber\\ 
\nabla_{a}F^{a b}&=0,\;\;\;\;\;\;\;T_{a b}=F_{d a}F^{d}_{b}-\dfrac{1}{4}g_{a b}F_{d e}F^{d e}
\end{align}
where $L$ is the Lagrangian of Einstein cubic gravity and $ P^{a b c d}=\partial L/\partial R_{a b c d} $ is
\begin{align}
P_{a b c d}&=g_{a [c}\;g_{b]d}+6\alpha[R_{a d}R_{b c}-R_{a c}R_{b d}+g_{bd}R_{a}^{e}R_{c e}-g_{a d}R_{b}^{e}R_{ce}-g_{b c}R_{a}^{e}R_{de}+g_{a c}R_{b}^{e}R_{de}\nonumber\\
&-g_{bd}R^{ef}R_{aecf}+g_{bc}R^{ef}R_{aedf}+g_{ad}R^{ef}R_{becf}-3R_{a}{}^{e}{}_{d}{}^{f}R_{becf}-g_{ac}R^{ef}R_{bedf}+3R_{a}{}^{e}{}_{c}{}^{f}R_{bedf}\nonumber\\
&+\dfrac{1}{2}R_{ab}{}^{ef}R_{cdef}].
\end{align}

Here, we consider the following spherically symmetric and static
line element for describing the geometry of spacetime
\begin{equation}\label{metform}
ds^{2}=-f(r)dt^{2}+\dfrac{dr^{2}}{f(r)}+r^2\left(d\theta^{2}+\dfrac{\sin^{2}(\sqrt{k}\theta)}{k}d\phi^{2}\right).
\end{equation}
As, we know generic static, spherically symmetric metrics do not need obey $g_{tt}g_{rr} = -1$ necessarily, but field equation (\ref{fieldeq3}) admits solutions with this property \cite{Bueno:2016lrh}-\cite{Frassino:2020zuv}, to which case we shall restrict our consideration in this section.

By inserting the metric into the field equations, the differential
equations for $ f(r)$ become
\begin{align}\label{eq6}
E^{r}_{r}-2T^{r}_{r}&=\dfrac{1}{4r^5}[24\alpha r^2ff^{\prime\prime\prime}(2k-2f+rf^{\prime})+24\alpha r^3ff^{\prime\prime 2}+r(r^4+12\alpha rff^{\prime}-96k\alpha f+96\alpha f^{2})f^{\prime\prime}\nonumber \\
&+24\alpha r(4f-k)f^{\prime 2}+4r^3(-k+f+\Lambda r^2)+(96k\alpha f+6r^4-96\alpha f^2)f^{\prime}]+\dfrac{q^2}{r^4}=0,
\end{align}
Expanding the function $f(r)$ around the event horizon $ r_{+} $
\begin{align}\label{eq7}
f(r)  &= f_{1}(r-r_{+})+f_{2}(r-r_{+})^{2}+f_{3}(r-r_{+})^{3}+...
\end{align}
and then inserting these expressions into equations (\ref{eq6}), we find
\begin{align}\label{eq9}
{f_{2}}&=\dfrac{-2\Lambda r_{+}^4+2kr_{+}^2+12\alpha kf_{1}^2-3f_{1}r_{+}^2-2q^2}{r_{+}^4},\nonumber\\
 {f_{3}}&=\dfrac{1}{3r_{+}^{5}+72\alpha r_{+}^3f_{1}^2+144\alpha r_{+}^2f_{1}k}[-6f_{2}r_{+}^4+f_{1}r_{+}^3(1-48\alpha f_{2}^2)+4r_{+}^2(24\alpha f_{1}^2f_{2}-k)\nonumber\\
 &+48\alpha f_{1}r_{+}(-f_{1}^2+3kf_{2})+8q^2-96\alpha kf_{1}^2]\nonumber,\\
 f_{4}&=-\dfrac{1}{r_{+}^{6}(96\alpha f_{1}k+48\alpha r_{+}f_{1}^2+r_{+}^3)}[20q^2+144\alpha kf_{2}f_{3}r_{+}^{3}+48\alpha f_{2}^{3}r_{+}^{4}-384\alpha f_{1}f_{2}^{3}r_{+}^{4}\nonumber\\
 &-384\alpha r_{+}^{3}f_{1}f_{2}^{2}+9r_{+}^{5}f_{3}-4r_{+}^{4}f_{2}-6kr_{+}^{6}-f_{1}r_{+}^{3}-360\alpha k f_{1}^{2}+624\alpha r_{+}^{2}f_{2}f_{1}^{2}-576\alpha r_{+}^{3}f_{1}^{2}f_{3}\nonumber\\
 &-240\alpha r_{+}f_{1}^{3}-792\alpha k r_{+}^{2}f_{1}f_{3}+504\alpha f_{1}f_{2}f_{3}r_{+}^{4}+720\alpha r_{+}k f_{1}f_{2}-144\alpha r_{+}^{2}k f_{2}^{2}]
\end{align}
where $r_+$, $f_1$ are undetermined constants of integration.
{In the large $r$ limit, we linearize the field equations about the Schwarzschild background
\begin{equation}
f(r)=1-\dfrac{2M}{r}+\epsilon F(r)
\end{equation}
where $F(r)$ should be calculated from the field equations. We linearize the differential equation by keeping terms only to order $\epsilon$, and the resulting differential equation for $F(r)$ takes the form
\begin{equation}\label{eqqdiff}
F^{\prime\prime\prime}+\gamma(r)F^{\prime\prime}+\eta(r)F^{\prime}+\omega(r)F+g(r)=0
\end{equation}
where
\begin{align}
\gamma(r)&=-\dfrac{96\alpha r^2+1152\alpha M^2-768\alpha Mr-96k\alpha r^2+r^6+192\alpha kMr}{48\alpha r(2M-r)(3M-r+kr)},\\
\eta(r)&=\dfrac{-16\alpha r^2 k+416\alpha M^2-240\alpha Mr+48\alpha kMr+16\alpha r^2-r^6}{8\alpha r^2(2M-r)(3M-r+kr)},\\
\omega(r)&=-\dfrac{288\alpha kMr-576\alpha Mr+1680\alpha M^2+r^6}{12\alpha r^3(2M-r)(3M+kr-r)},\\
g(r)&=\dfrac{(k-1)r^7-q^2r^5-288M\alpha r^2(k-1)+120\alpha M^2 r(5k-14)+2208\alpha M^3}{12\alpha r^4(2M-r)(3M-r+kr)}.
\end{align}
In the large $r$ limit and $k=1$, the homogenous equation reads
\begin{equation}
F^{\prime\prime\prime}+\dfrac{r^4}{144M\alpha}F^{\prime\prime}+\dfrac{r^3}{24M\alpha}F^{\prime}+\dfrac{r^2}{36M\alpha}F=0.
\end{equation}
This equation can be solved exactly in terms of hypergeom functions
\begin{align}
F(r)&=c_{1}\text{hypergeom}\left(\left[\dfrac{1}{5}\right],\left[\dfrac{3}{5}\right],-\dfrac{r^5}{720M\alpha}\right)+c_{2}r^2\text{hypergeom}\left(\left[\dfrac{3}{5}\right],\left[\dfrac{7}{5}\right],-\dfrac{r^5}{720M\alpha}\right)\nonumber\\
&+c_{3}r\text{hypergeom}\left(\left[\dfrac{2}{5},1\right],\left[\dfrac{4}{5},\dfrac{6}{5}\right],-\dfrac{r^5}{720M\alpha}\right).
\end{align}
In tha large $r$, $F(r)$ decays super-exponentially and can be neglected. 
More relevant is the particular solution, which reads
\begin{align}\label{eqqqhfo}
 f_{p}(r)=&\sum_{n=2}\dfrac{F_{n}}{r^{n}}=\dfrac{F_{2}}{r^2}+\dfrac{F_{3}}{r^3}+....
\end{align}
By inserting the above expansions into the field equations  (\ref{eqqdiff}) and solving order by order, one can get
\begin{align}\label{eqasymp}
f_{p}(r)&=\dfrac{q^2}{r^2}+\dfrac{F_{4}}{r^4}-\dfrac{432\alpha M^{2}}{r^6}+\dfrac{64M\alpha(23M^{2}+18q^2)}{3r^{7}}-\dfrac{4656M^2q^2\alpha}{7r^{8}}+\dfrac{432M\alpha F_{4}}{r^{9}}\nonumber\\
&-\dfrac{736M^{2}F_{4}\alpha}{r^{10}}-\dfrac{995328M^{3}\alpha^2}{5r^{11}}+\dfrac{3456M^{2}\alpha^{2}(1831M^{2}+576q^{2})}{11r^{12}}+\mathcal{O}\left(\dfrac{1}{r^{13}}\right).
\end{align}
For $\alpha$ goes to zero, one expects the metric returning to the RN. As a result $F_{4}$ should be zero. The solution at large $r$, thereby giving
\begin{align}\label{eqqasymp}
f(r)&\approx 1-\dfrac{2M}{r}+\dfrac{q^2}{r^2}-\dfrac{432\alpha M^{2}}{r^6}+\dfrac{64M\alpha(23M^{2}+18q^2)}{3r^{7}}-\dfrac{4656M^2q^2\alpha}{7r^{8}}-\dfrac{995328M^{3}\alpha^2}{5r^{11}}\nonumber\\
&+\dfrac{3456M^{2}\alpha^{2}(1831M^{2}+576q^{2})}{11r^{12}}.
\end{align}
 }
It should be noted, the other components of field equations give the same results for (\ref{eq7})-(\ref{eqasymp}).
We wish to obtain an approximate analytic solution (for $k=1$) that is valid
near the horizon and at large $r$. To this end we employ a
continued fraction expansion \cite{Rezzolla:2014mua}, and write
\begin{equation}\label{eq17}
f(r)=xA(x),\hspace{1cm} x= 1- \frac{r_+}{r}
\end{equation}
with
\begin{align}
A(x) &=1-\epsilon(1-x)+(a_{0}-\epsilon)(1-x)^{2}+\dfrac{a_{1}(1-x)^{3}}{1+\dfrac{a_{2}x}{1+\dfrac{a_{3}x}{1+\dfrac{a_{4}x}{1+...}}}}
\label{Ax}
\end{align}
where we truncate the continued fraction at order $4$.
By expanding (\ref{eq17}) near the horizon ($ x\to 0 $) and
the asymptotic  region ($ x\to 1 $)  we obtain
\be
\epsilon=-\dfrac{F_{1}}{r_{+}}-1, \qquad a_{0}=\dfrac{q^2}{r_{+}^2},\qquad a_{1}=-1-a_{0}+2\epsilon+r_{+}f_{1}
\ee
for the lowest order expansion coefficients, with the remaining
$a_i$ given in terms of $(r_+, f_1)$; we provide these expressions in the  Appendix \ref{appa}.

The result is an approximate analytic solution for metric functions everywhere outside the horizon. For a static space time
we have a timelike Killing vector
$ \xi=\partial_{t} $ everywhere outside the horizon and so  we obtain
\begin{align}\label{eq20}
T &=\left. \dfrac{f^{'}(r)}{4\pi}\right\vert_{r_{+}}
= \dfrac{f_{1}}{4\pi} = {\dfrac {(1-2\epsilon+a_{1}+a_{0})}{{4\pi r_{+}}}}=\dfrac{1+\delta(r_{+},q)}{4\pi r_{+}}.
 \end{align}
 Extreme charged black hole solutions exist if $f_{1}=0$ implying that $a_{1}=2\epsilon-a_{0}-1$.
We compute the entropy as follows \cite{Wald1}, \cite{Wald2}
\begin{align}\label{eqqqentropy}
S=-2\pi\int_{Horizon}d^{2}x\sqrt{\eta}\dfrac{\delta L}{\delta R_{a b c d}}\epsilon_{a b}\epsilon_{c d}&=\dfrac{A}{4}\left[1+6\alpha \left(\dfrac{f^{\prime 2}}{r_{+}^2}+\dfrac{4f^{\prime}}{r_{+}^3}\right)\right]
=\dfrac{A}{4}\left[1+6\alpha \left(\dfrac{f^{2}_{1}}{r_{+}^2}+\dfrac{4f_{1}}{r_{+}^3}\right)\right]\nonumber\\
&=\dfrac{A}{4}\left[1+\dfrac{6\alpha(1+\delta(r_{+},q))}{r_{+}^{4}}[5+\delta(r_{+},q)]\right].
\end{align}
We now consider the thermodynamics of these black hole solutions,
whose basic equations are the first law and Smarr formula
\begin{equation}\label{eqfirstlaw}
dM=TdS+\phi dq,
\end{equation}
\begin{equation}\label{eq26}
M=2TS+q\phi,
\end{equation}
where there are no pressure/volume terms since we have set $\Lambda=0$.
From Eq. \eqref{eq26} we have
\begin{equation}\label{eqmass}
M=\dfrac{r_{+}^{4}+30\alpha+66\alpha\delta(r_{+},q)+42\alpha\delta^{2}(r_{+},q)
+r_{+}^{4}\delta(r_{+},q)+6\alpha\delta^{3}(r_{+},q)+2q^2r_{+}^2}{2r_{+}^2},
\end{equation}
yielding the mass parameter as a function of the horizon radius. In the following, we show that the asymptotic behavior of the mass \eqref{eqmass} is the same as the mass of the Schwarzschild black hole. So, one can interpret the mass \eqref{eqmass}, as ADM mass. 
 We now impose the first law \eqref{eqfirstlaw}, which becomes
\begin{equation}\label{flaw1}
\dfrac{\partial M}{\partial r_{+}}dr_{+}+\dfrac{\partial M}{\partial q}dq=T \dfrac{\partial S}{\partial r_{+}}dr_{+}+T \dfrac{\partial S}{\partial q}dq+\phi   dq
\end{equation}
yielding
\begin{align}\label{flawrp}
&\dfrac{\partial M}{\partial r_{+}}-T\dfrac{\partial S}{\partial r_{+}}=0,\hspace{0.5cm}\Longrightarrow\nonumber\\
&60\alpha +132\alpha\delta(r_{+},q)+84\alpha \delta^{2}(r_{+},q)+12\alpha\delta^{3}(r_{+},a)+2q^2r_{+}^{2}-48\alpha r_{+}\dfrac{\partial\delta(r_{+},q)}{\partial r_{+}}\nonumber\\
&-60\alpha r_{+}\delta(r_{+},q)\dfrac{\partial\delta(r_{+},q)}{\partial r_{+}}-r_{+}^{5}\dfrac{\partial\delta(r_{+},q)}{\partial r_{+}}-12\alpha r_{+}\delta^{2}(r_{+},q)\dfrac{\partial\delta(r_{+},q)}{\partial r_{+}}=0
\end{align}
and
\begin{align}\label{flawq}
&\dfrac{\partial M}{\partial q}-T\dfrac{\partial S}{\partial q}-\phi=0,\hspace{0.5cm}\Longrightarrow\nonumber\\
&48\alpha \dfrac{\partial\delta(r_{+},q)}{\partial q}+60\alpha \delta(r_{+},q)\dfrac{\partial\delta(r_{+},q)}{\partial q}+r_{+}^{4}\dfrac{\partial\delta(r_{+},q)}{\partial q}+12\alpha \delta^{2}(r_{+},q)\dfrac{\partial\delta(r_{+},q)}{\partial q}+2q^2 r_{+}^{2}=0
\end{align}
as differential equations that must be satisified by $\delta(r_{+},q)$. From now, we consider the case $q=0$.
So, from equation (\ref{flawrp}) one can get
\begin{equation}
r_{+}^{4}(1+\delta(r_{+}))^2+24\alpha (1+\delta(r_{+}))^3+6\alpha(1+\delta(r_{+}))^4+4c_{1}r_{+}^{4}=0,
\end{equation}
by solving above equation one can obtain solutions for $\delta(r_{+})$ as:
\begin{align}
\delta^{1,2}&=-2+\dfrac{\sqrt{2\mathcal{U}}}{12\sqrt{\alpha}\mathcal{Q}^{\frac{1}{6}}}\pm\nonumber\\
&\dfrac{i\sqrt{2}}{\sqrt{\alpha}\mathcal{Q}^{\frac{1}{6}}\mathcal{U}^{\frac{1}{4}}}\sqrt{864\sqrt{2\mathcal{Q}}\alpha^{\frac{3}{2}}+\sqrt{\mathcal{U}}\mathcal{Q}^{\frac{2}{3}}r_{+}+\sqrt{\mathcal{U}}(r_{+}^7+288c_{1}\alpha r_{+}^3+\mathcal{Q}^{\frac{1}{3}}(4r_{+}^4-144\alpha))-36\sqrt{2\alpha \mathcal{Q}}r_{+}^4}\label{eqq2223}\\
\delta^{3,4}&=-2+\dfrac{\sqrt{2\mathcal{U}}}{12\sqrt{\alpha}\mathcal{Q}^{\frac{1}{6}}}\pm\nonumber\\
&\dfrac{\sqrt{2}}{\sqrt{\alpha}\mathcal{Q}^{\frac{1}{6}}\mathcal{U}^{\frac{1}{4}}}\sqrt{864\sqrt{2\mathcal{Q}}\alpha^{\frac{3}{2}}-\sqrt{\mathcal{U}}\mathcal{Q}^{\frac{2}{3}}r_{+}-\sqrt{\mathcal{U}}(r_{+}^7+288c_{1}\alpha r_{+}^3+\mathcal{Q}^{\frac{1}{3}}(4r_{+}^4-144\alpha))-36\sqrt{2\alpha \mathcal{Q}}r_{+}^4}\label{eqq2225}
\end{align}
where
\begin{align}\label{eqq2226}
\mathcal{P}&=36\sqrt{2\alpha c_{1}}\sqrt{192c_{1}\alpha r_{+}^8+373248c_{1}\alpha^3 +24\alpha r_{+}^8-9216\alpha^2 c_{1}^2r_{+}^4-20736c_{1}\alpha^2 r_{+}^4}\nonumber\\
\mathcal{Q}&=(-864c_{1}r_{+}^4\alpha+31104c_{1}\alpha^2 +r_{+}^8+\mathcal{P})r_{+}\nonumber\\
\mathcal{U}&=72\alpha \mathcal{Q}^{\frac{1}{3}}-2r_{+}^4\mathcal{Q}^{\frac{1}{3}}+\mathcal{Q}^{\frac{2}{3}}r_{+}+288r_{+}^3c_{1}\alpha +r_{+}^7.
\end{align}

\begin{figure}[H]\hspace{0.4cm}
\centering
\subfigure{\includegraphics[width=0.3\columnwidth]{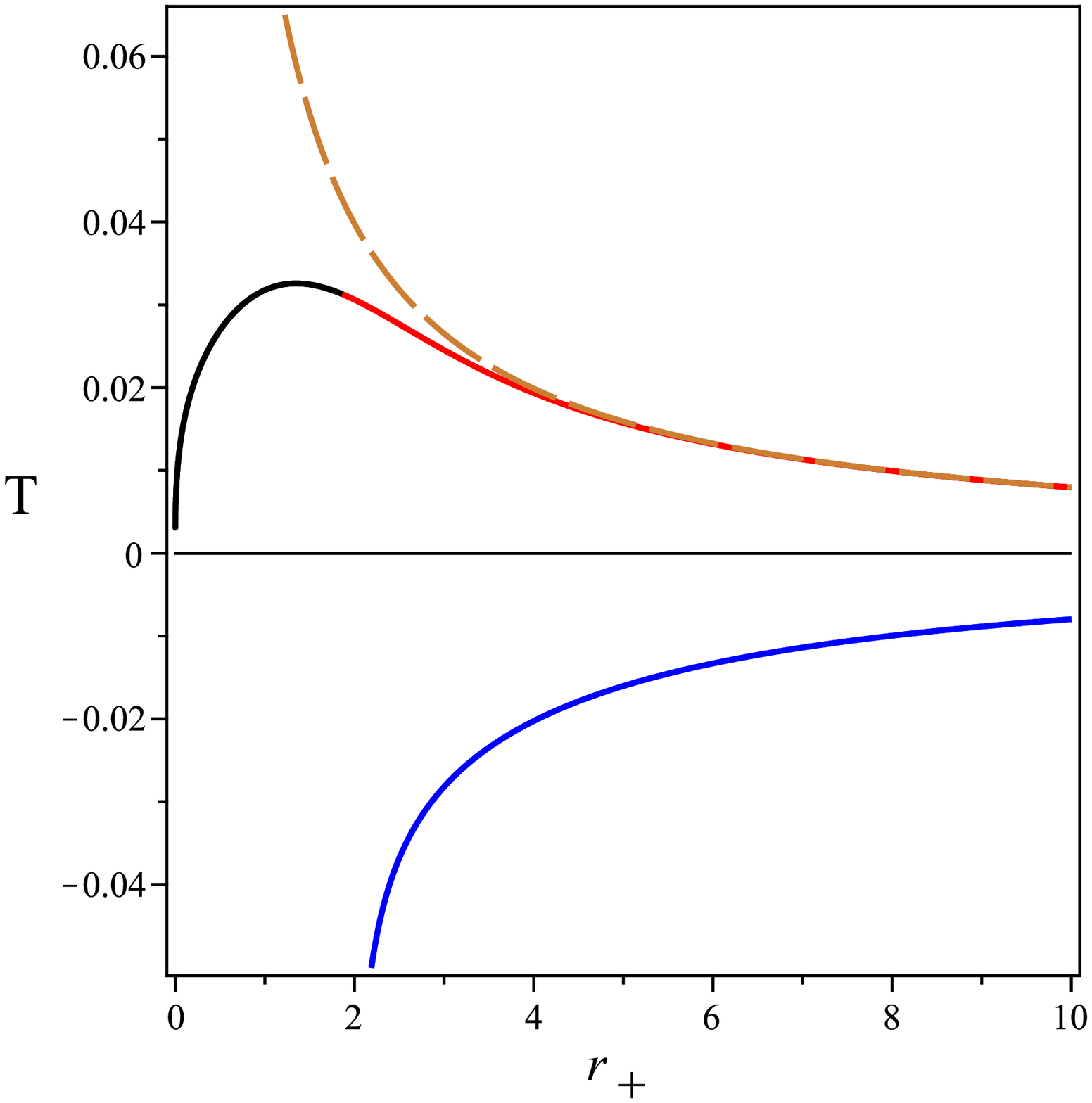}}
\subfigure{\includegraphics[width=0.3\columnwidth]{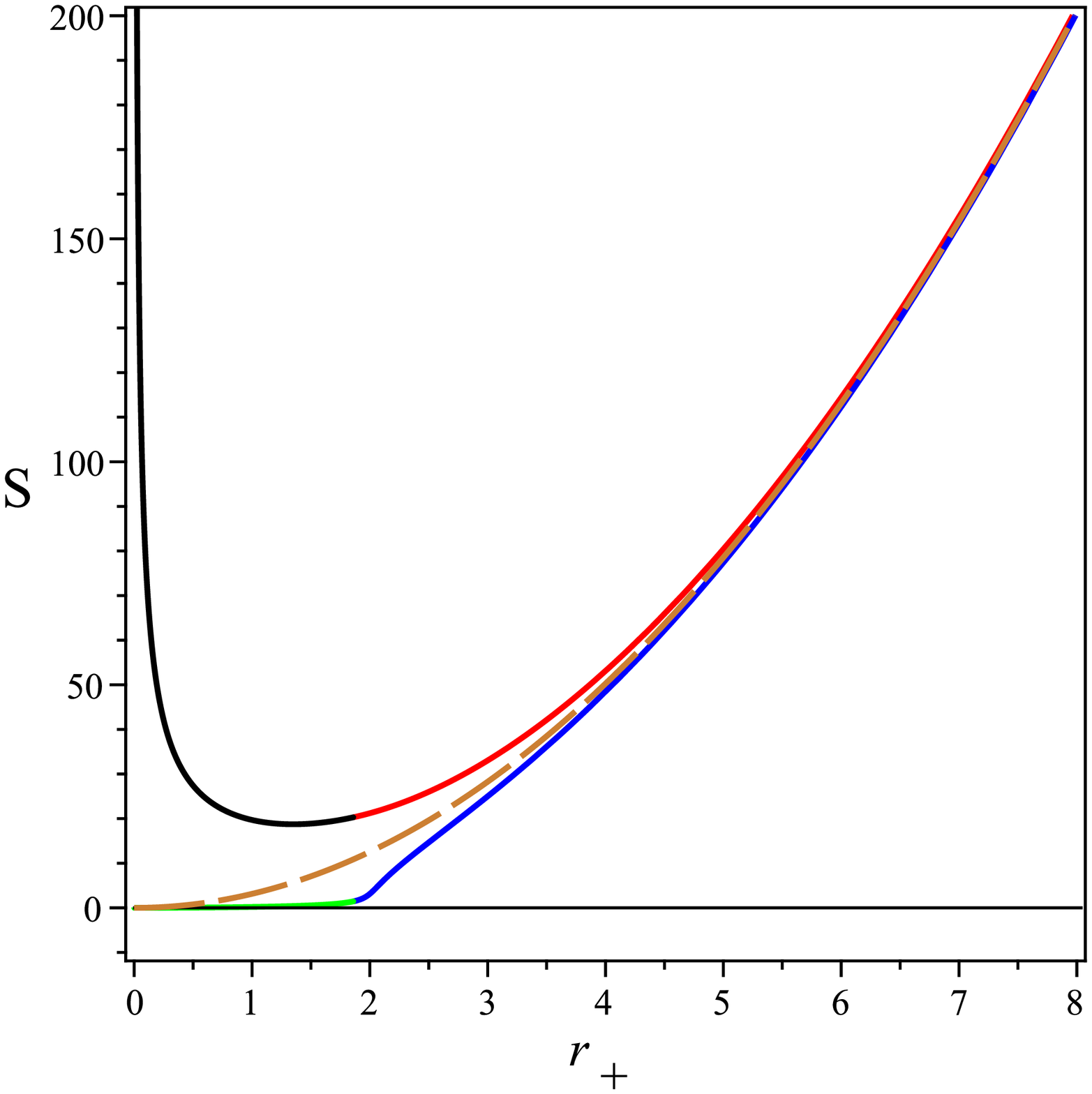}}
\subfigure{\includegraphics[width=0.3\columnwidth]{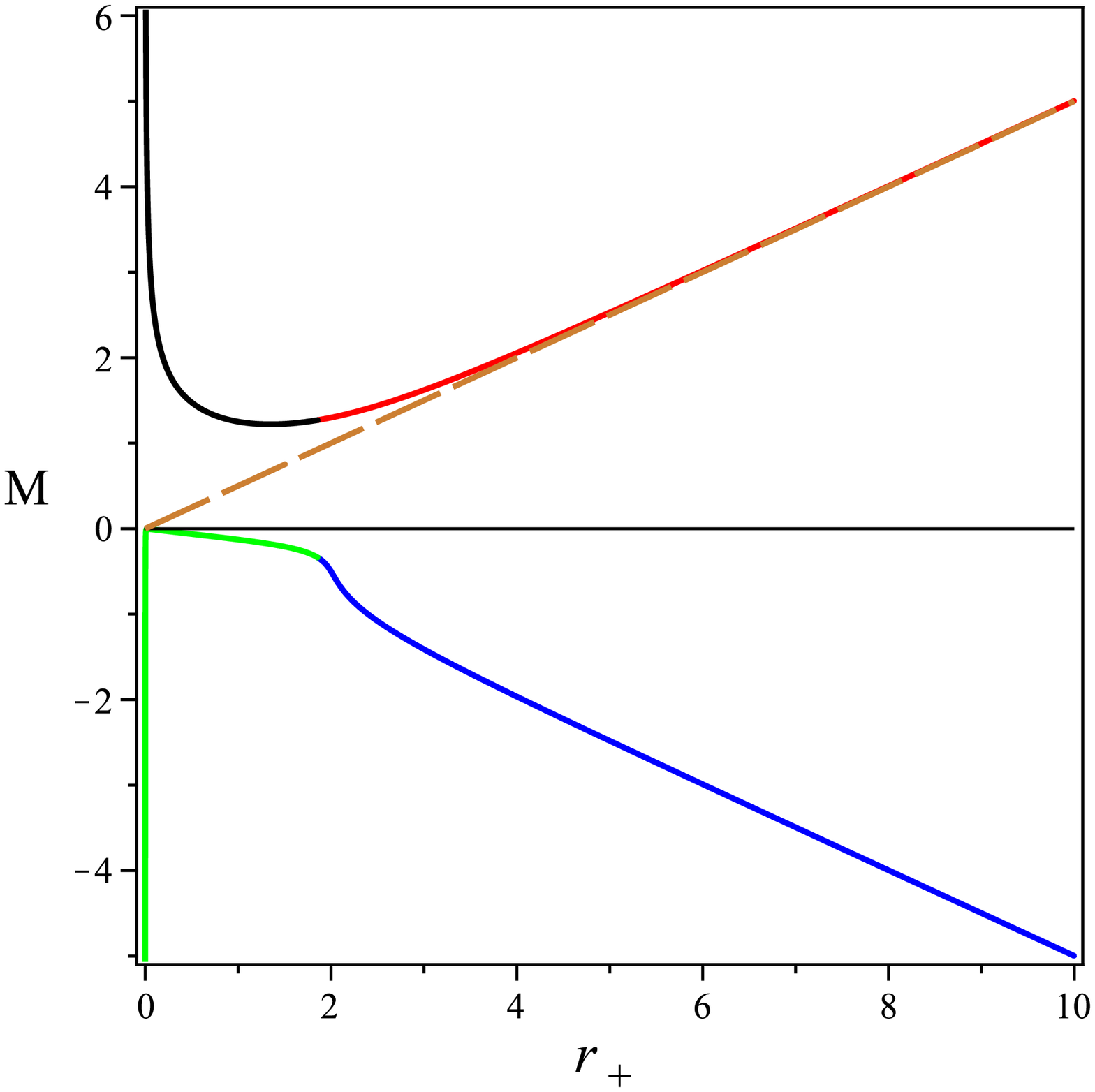}}
\caption{The temperature, entropy and mass as a function of
$r_{+}$ for $c_{1}=-0.25, \alpha=0.5$. The orange
long dashed lines related to the Schwarzschild’s black hole.} \label{TT-Mmplot}
\end{figure}

\begin{figure}[H]\hspace{0.4cm}
\centering
\subfigure{\includegraphics[width=0.4\columnwidth]{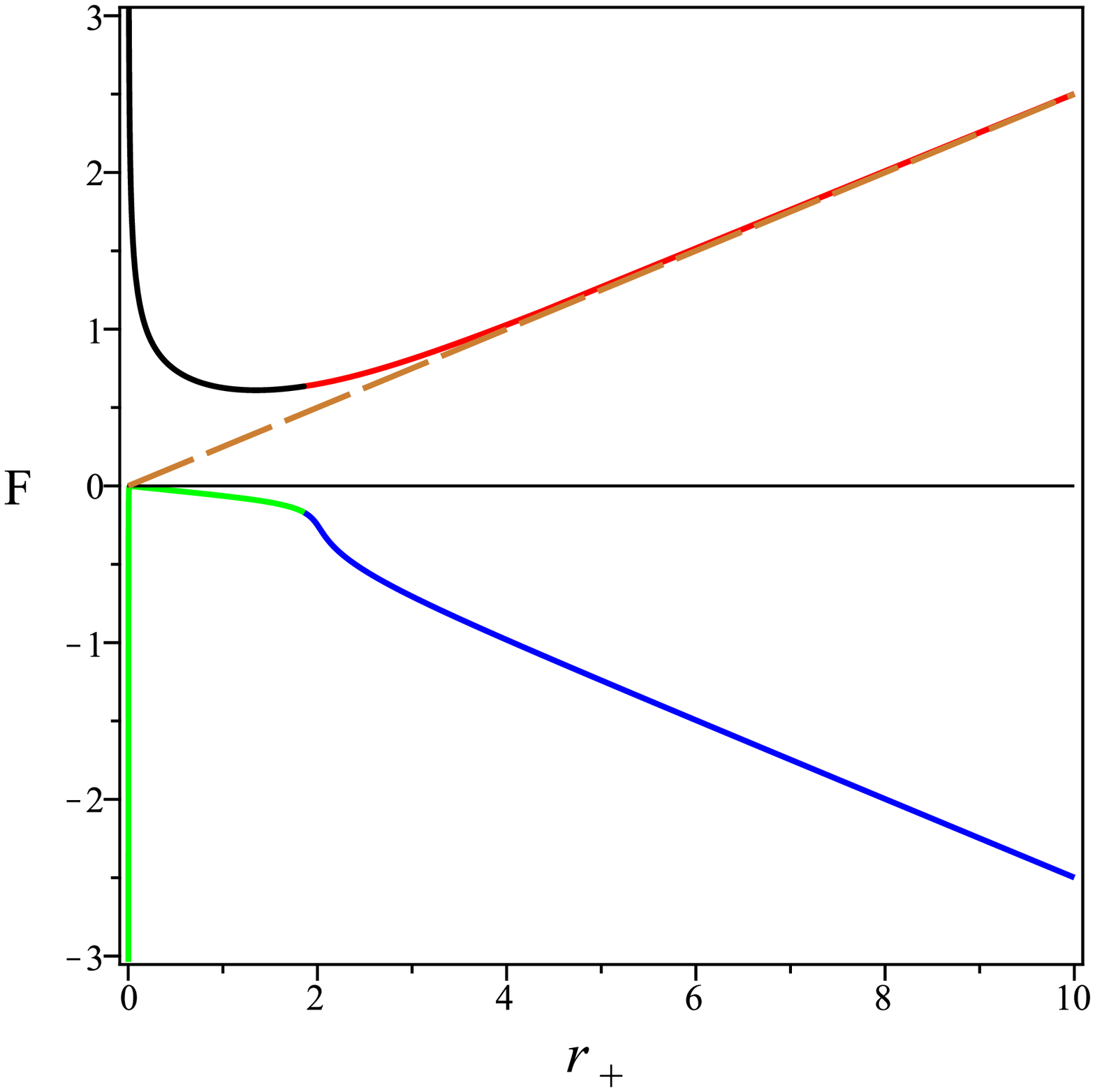}}
\subfigure{\includegraphics[width=0.4\columnwidth]{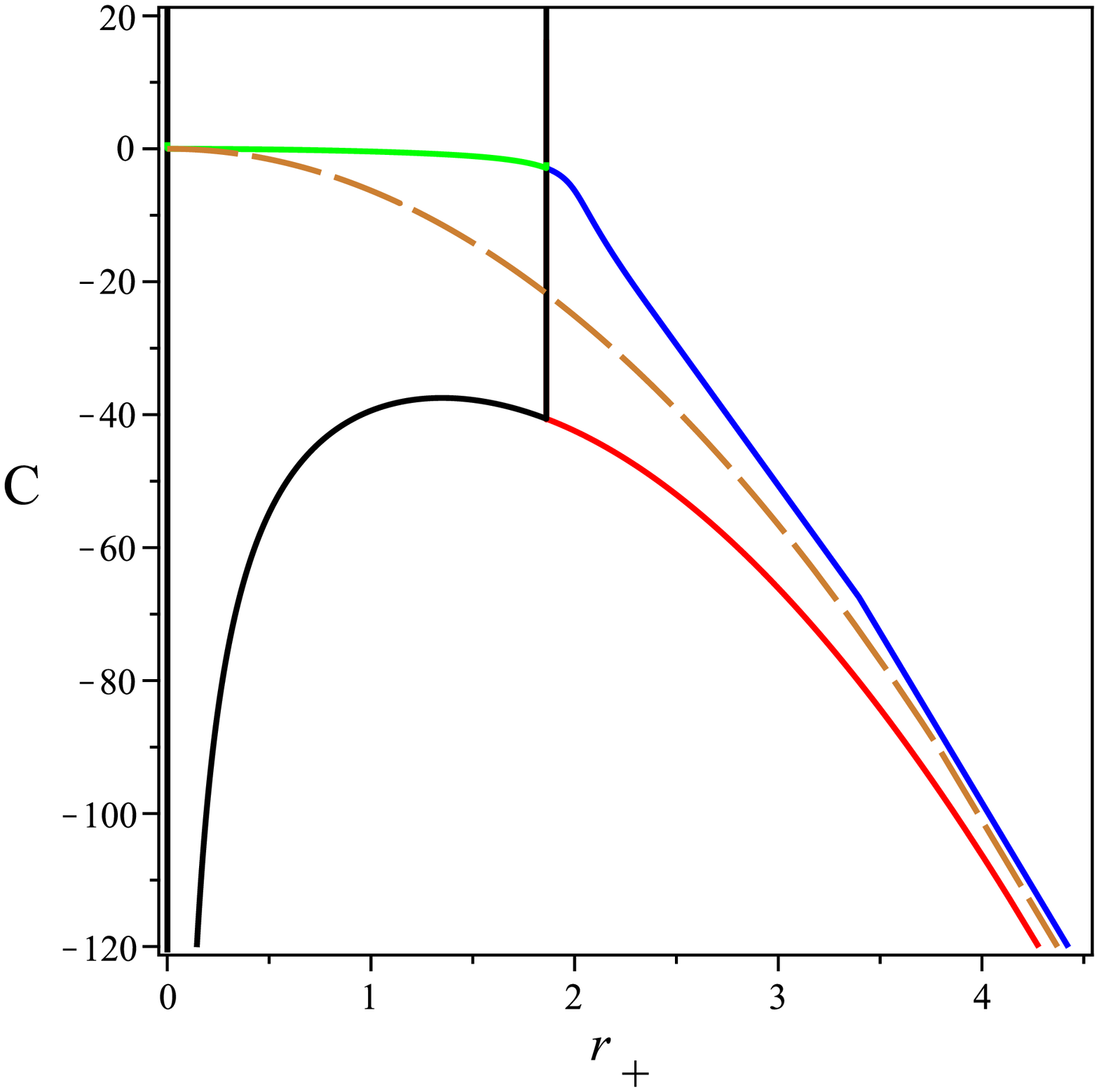}}
\caption{The free energy and heat capacity as a function of $r_{+}$ for $c_{1}=-0.25, \alpha=0.5$. The orange long dashed lines related to the Schwarzschild’s black hole.}
\label{Cplot0}
\end{figure}

\begin{figure}[H]\hspace{0.4cm}
\centering
\subfigure{\includegraphics[width=0.3\columnwidth]{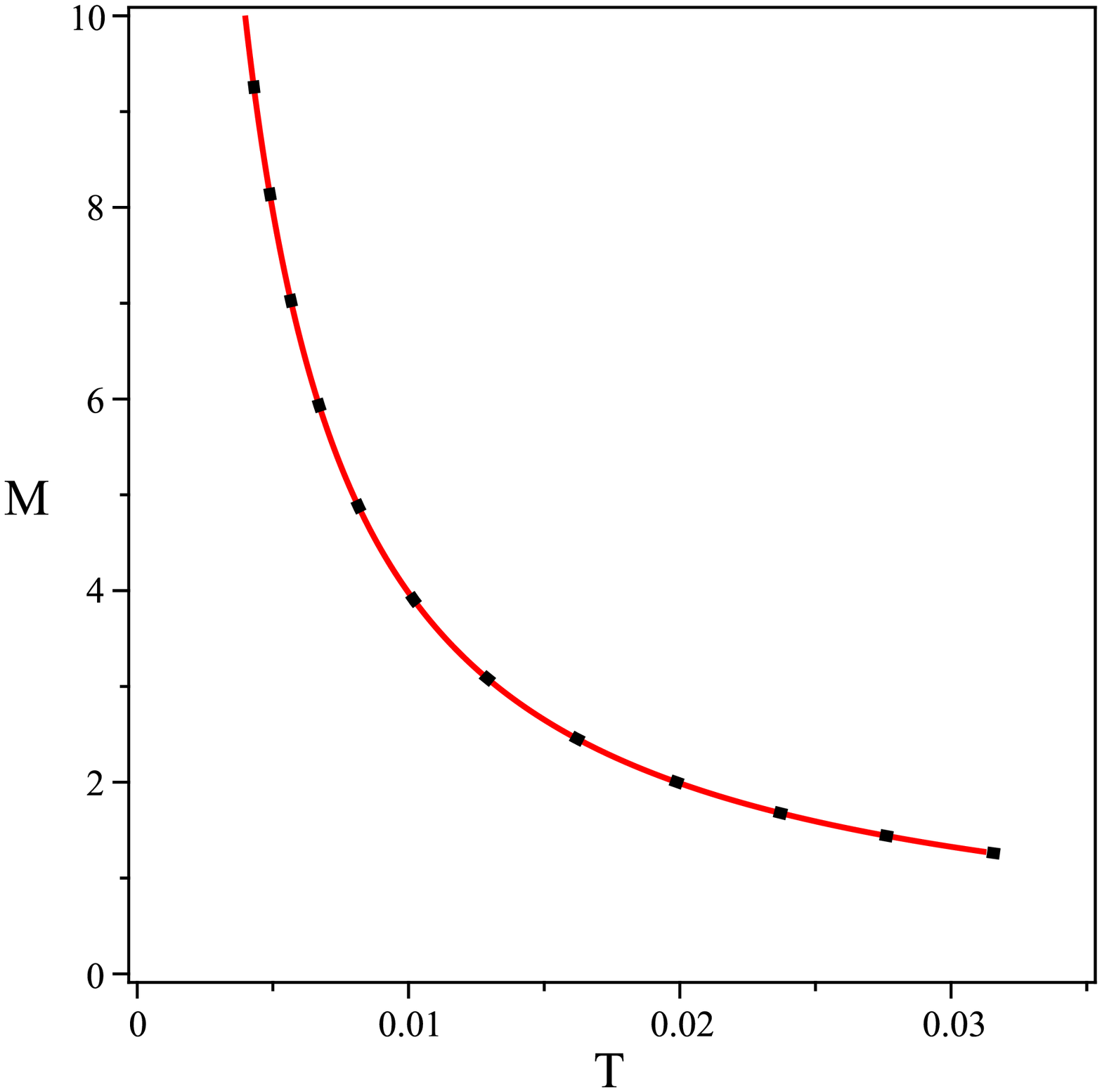}}
\subfigure{\includegraphics[width=0.3\columnwidth]{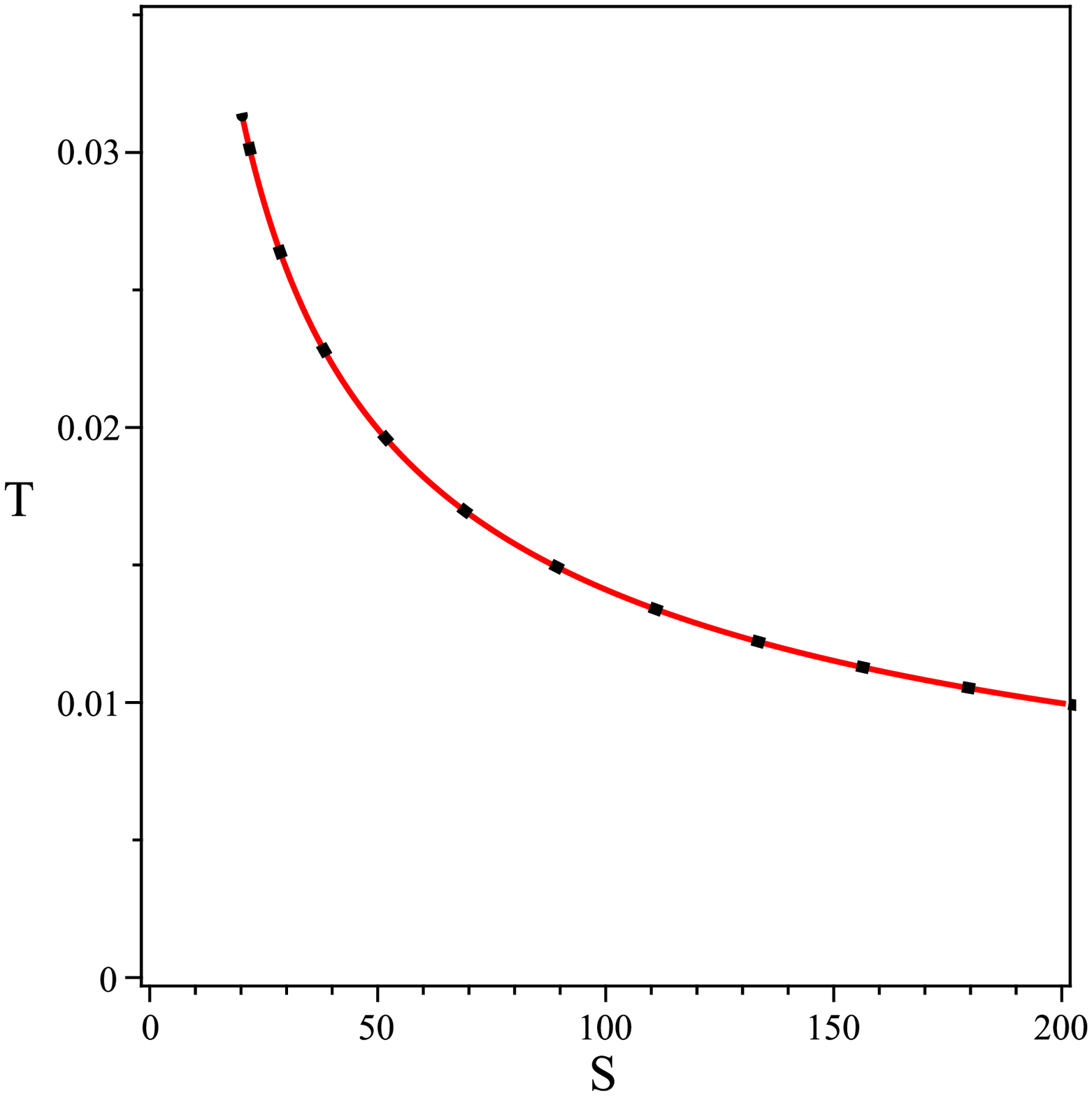}}
\subfigure{\includegraphics[width=0.3\columnwidth]{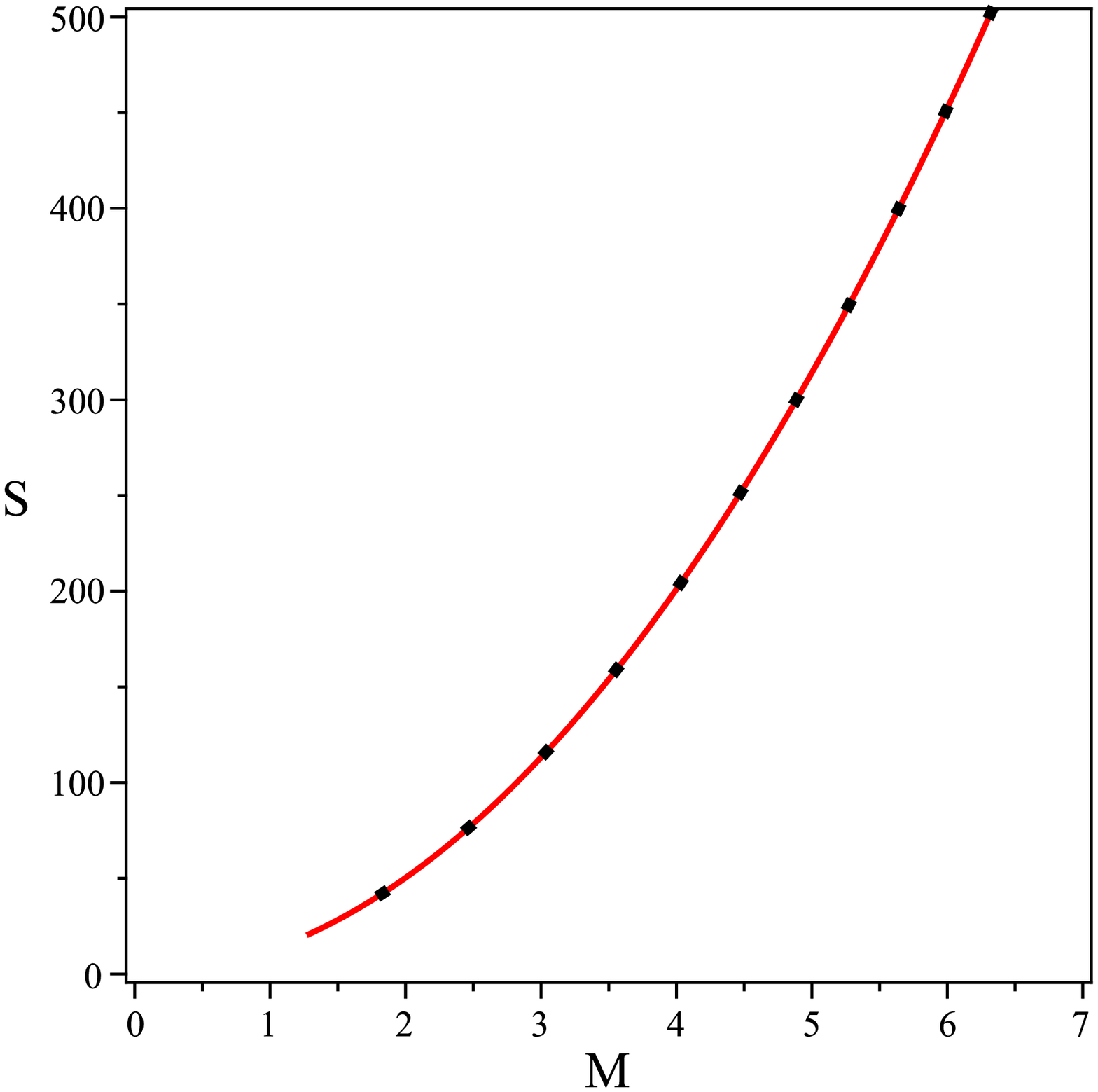}}
\caption{The plots of temperature, entropy and mass for
$c_{1}=-0.25, \alpha=0.5$.} \label{Ss-Mmplot0}
\end{figure}

Inserting \eqref{eqq2223}-\eqref{eqq2226} into the thermodynamical
quantities and plotting them, one can arrive Figs.
\ref{TT-Mmplot}, \ref{Cplot0} and \ref{Ss-Mmplot0}. In Fig.
\ref{TT-Mmplot}, we have illustrated the temperature, entropy, and
mass for different values of parameters. In these figures, the
black and red solid lines correspond to the physical solutions,
while the green and blue solid lines correspond to the
non-physical black hole solutions. The physical solutions have
positive temperatures and mass. In these figures, the behavior of
the red solid lines is similar to the Schwarzschild black hole,
and the behavior of the black solid lines is different from that
of Schwarzschild black holes.  In Fig. \ref{TT-Mmplot}a,  we
observe that, unlike the Einstein gravity case, it no longer
diverges for $r_{+}\to 0$. Instead, there is a maximum value
$T_{max}$ that is reached at $r_{+,max}$.

\begin{figure}[H]\hspace{0.4cm}
\centering
\subfigure{\includegraphics[width=0.3\columnwidth]{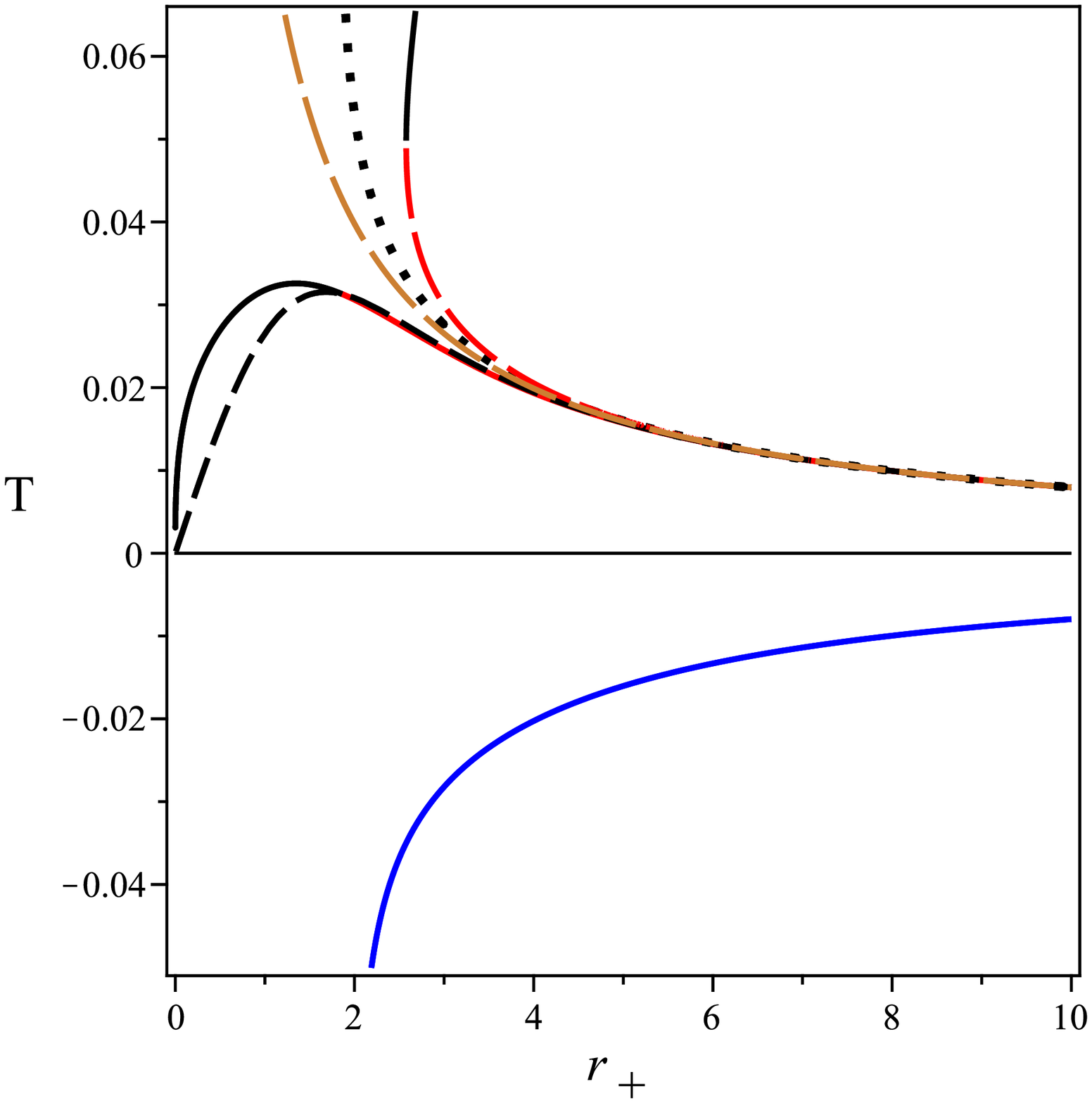}}
\subfigure{\includegraphics[width=0.3\columnwidth]{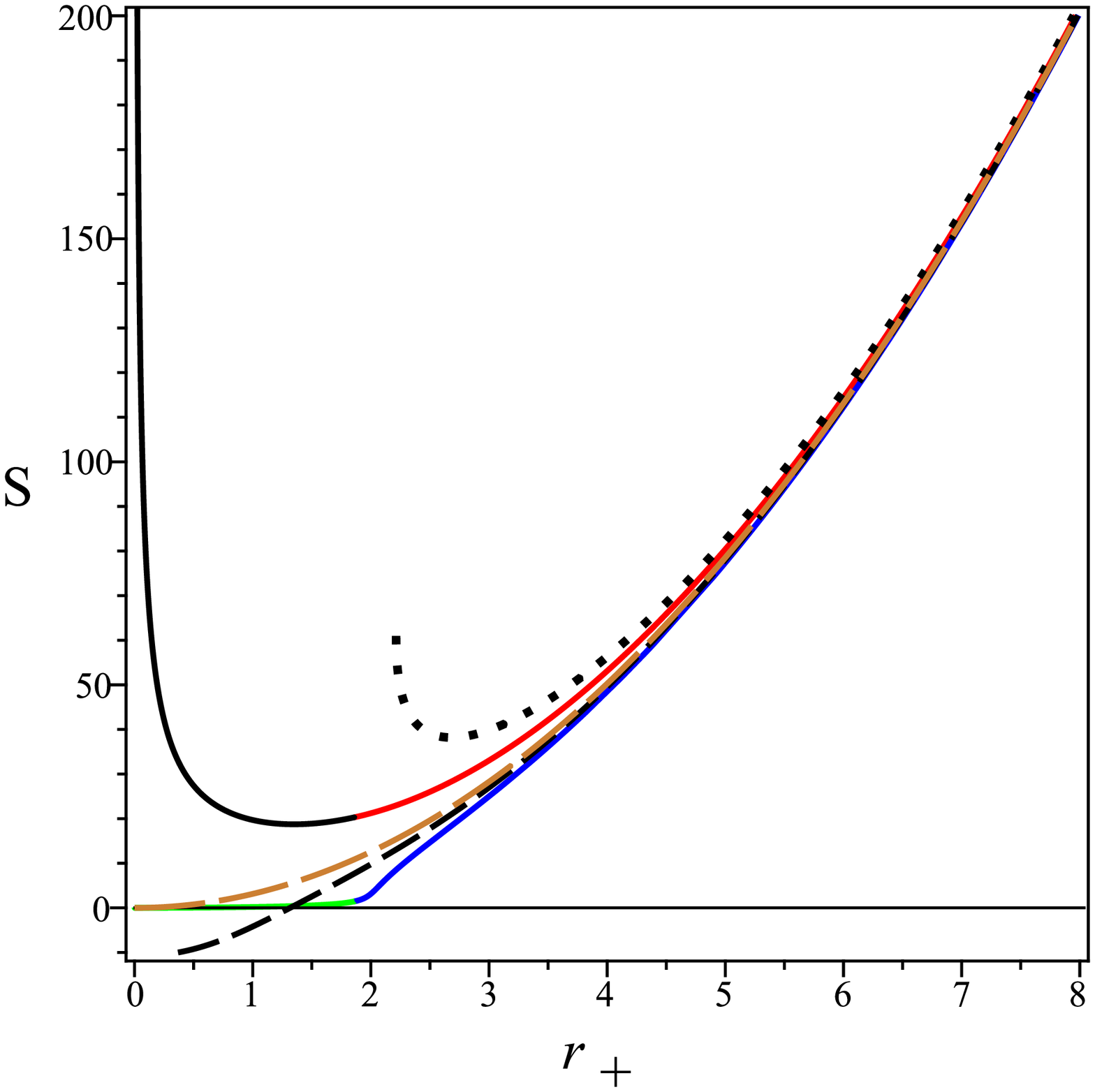}}
\subfigure{\includegraphics[width=0.3\columnwidth]{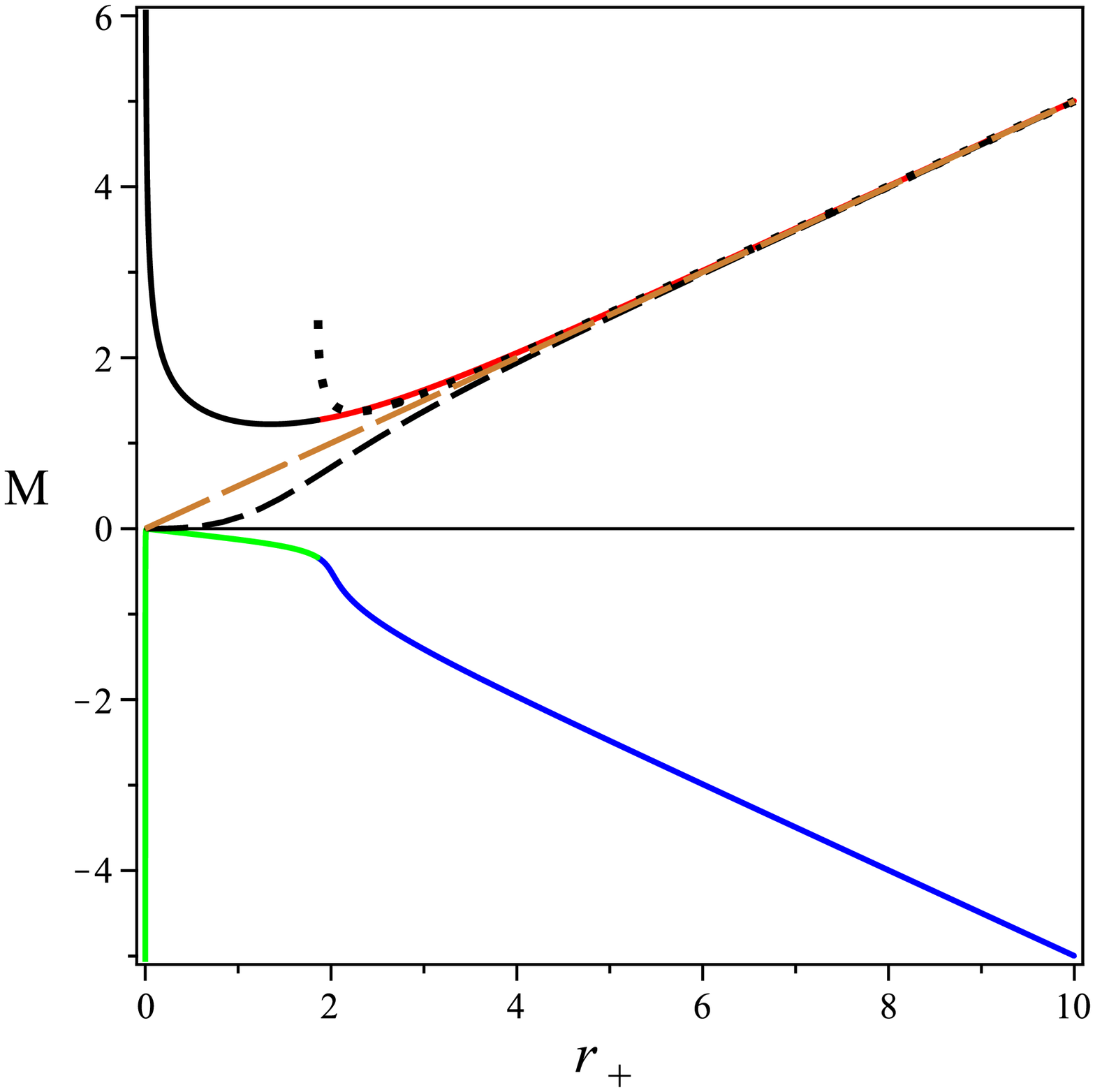}}
\caption{The temperature, entropy and mass as a function of
$r_{+}$ for $c_{1}=-0.25, \alpha=0.5$: solid lines are the results
of our method and dashed lines are the results of the paper
\cite{Bueno:2016lrh}-\cite{Hennigar:2018hza}. For $\alpha=-0.5$:
Long dashed lines are the results of our method and  dotted lines
are for the paper \cite{Bueno:2016lrh}-\cite{Hennigar:2018hza}. The orange
long dashed lines related to the Schwarzschild’s black hole.}
\label{CplotC}
\end{figure}

In Fig. \ref{TT-Mmplot}b, c, the behavior of entropy and mass in
the large radii are similar to the Schwarzschild black hole and in
the small radii, unlike the Einstein gravity case, diverge for
$r_{+}\to 0$. Different panels of Fig. \ref{CplotC}, compare our
black hole solutions (solid and long dashed lines) with those
obtained in \cite{Bueno:2016lrh,Hennigar:2018hza,Hennigar:2016gkm}
(dashed and dotted lines). The solid and dashed lines are the
results of our calculus and the paper
\cite{Bueno:2016lrh,Hennigar:2018hza,Hennigar:2016gkm} for
positive $\alpha$, respectively. The long dashed and dotted lines
are the results of our calculus and the paper
\cite{Bueno:2016lrh,Hennigar:2018hza,Hennigar:2016gkm} for
negative $\alpha$, respectively. As can be seen, there is a good
agreement in both cases between our computations and the results
of the mentioned papers in large radii and somewhat different in
small radii but have the same behavior. This shows that the first law of thermodynamic and the Smarr formula in the form of (\ref{eqfirstlaw}) and (\ref{eq26}) are valid for this theory of gravity.

In the following, we are going to look at the thermodynamical stability of the solutions.
In global stability, we allow a system in equilibrium with a
thermodynamic reservoir to exchange energy with the reservoir. The
preferred phase of the system is the one that minimizes the free
energy. In order to investigate the global stability, we use the
following expression for the free energy
\begin{equation}
F=M-TS.
\end{equation}
In Fig. \ref{Cplot0}a, we have shown the free energy in terms of
$r_{+}$ for positive $\alpha$. As can be seen in the black, green,
and blue branches the free energy is decreasing functions of
$r_{+}$. This shows the black holes in these branches globally are
stable. While in the red branch, the free energy has an increasing
behavior and is a globally unstable branch.

On the other hand, local stability is concerned with how the
system responds to small changes in its thermodynamic parameters.
In order to study the thermodynamic stability of the black holes
with respect to small variations of the thermodynamic coordinates,
one can investigate the behavior of the heat capacity. The
positivity of the heat capacity ensures local stability. The heat
capacity is given by
\begin{equation}
C=\dfrac{\partial M}{\partial T}.
\end{equation}
In Fig. \ref{Cplot0}b, heat capacity in terms of $r_{+}$ for
positive $\alpha$ have been illustrated. As can be seen the heat
capacity in all radii is negative, this shows the black hole in
all branches locally unstable.

\begin{figure}[H]\hspace{0.4cm}
\centering
\subfigure{\includegraphics[width=0.4\columnwidth]{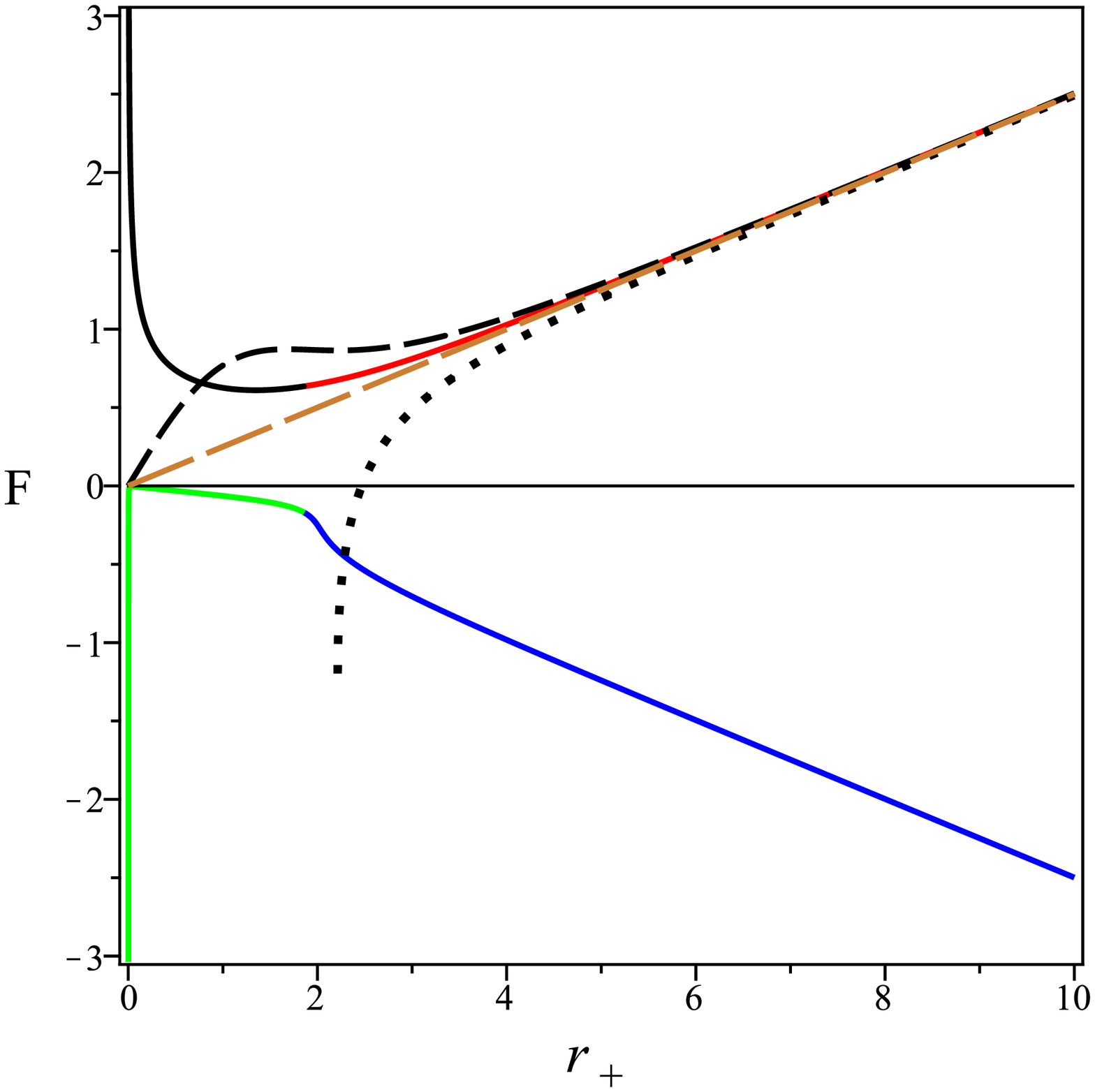}}
\subfigure{\includegraphics[width=0.4\columnwidth]{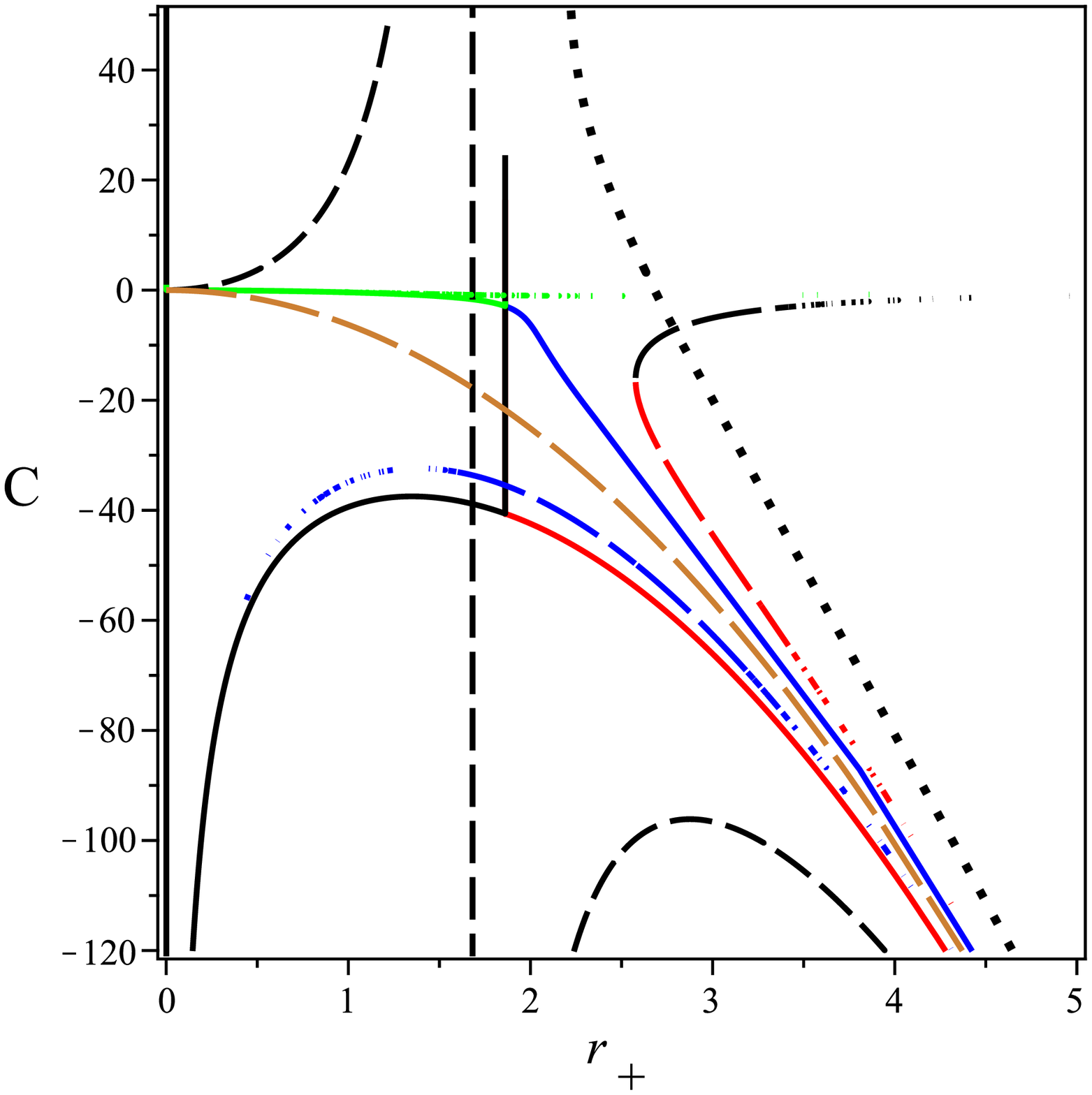}}
\caption{The free energy and heat capacity as a function of
$r_{+}$ for $c_{1}=-0.25, \alpha=0.5$: solid lines are the results
of our method and dashed lines are the results of the paper
\cite{Bueno:2016lrh}-\cite{Hennigar:2018hza}. For $\alpha=-0.5$: Long
dashed lines are the results of our method and dotted lines are
for the paper \cite{Bueno:2016lrh}-\cite{Hennigar:2018hza}. The orange
long dashed lines related to the Schwarzschild’s black hole.}
\label{Cplot}
\end{figure}

In Fig. \ref{Cplot}, the heat capacity and free energy are
compared based on our calculations and the results of papers
\cite{Bueno:2016lrh}-\cite{Hennigar:2018hza}. As can be seen from
the dashed lines of both figures, the black holes in small radii
are locally stable but globally unstable.

\begin{figure}[H]\hspace{0.4cm}
\centering
\subfigure{\includegraphics[width=0.3\columnwidth]{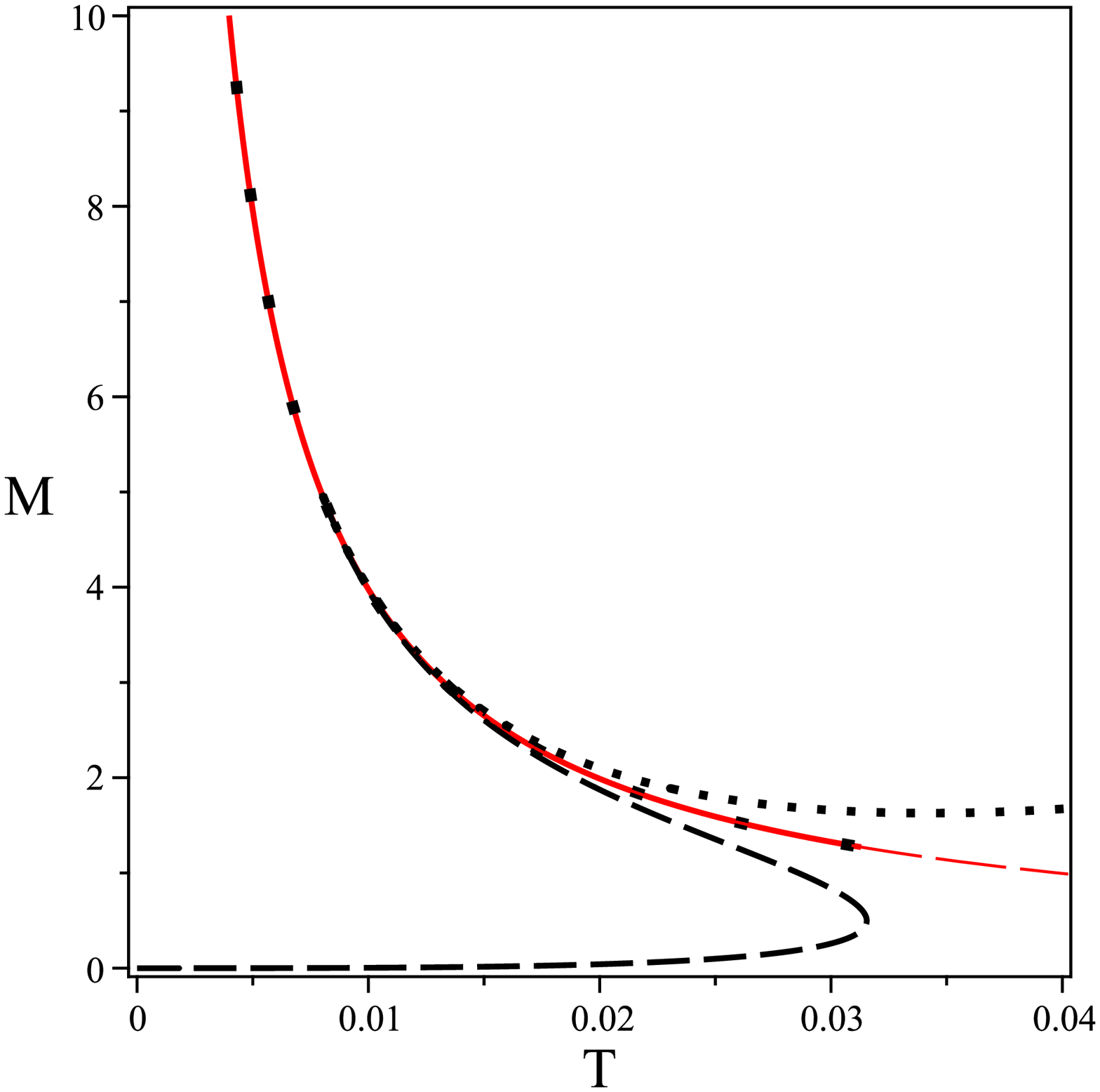}}
\subfigure{\includegraphics[width=0.3\columnwidth]{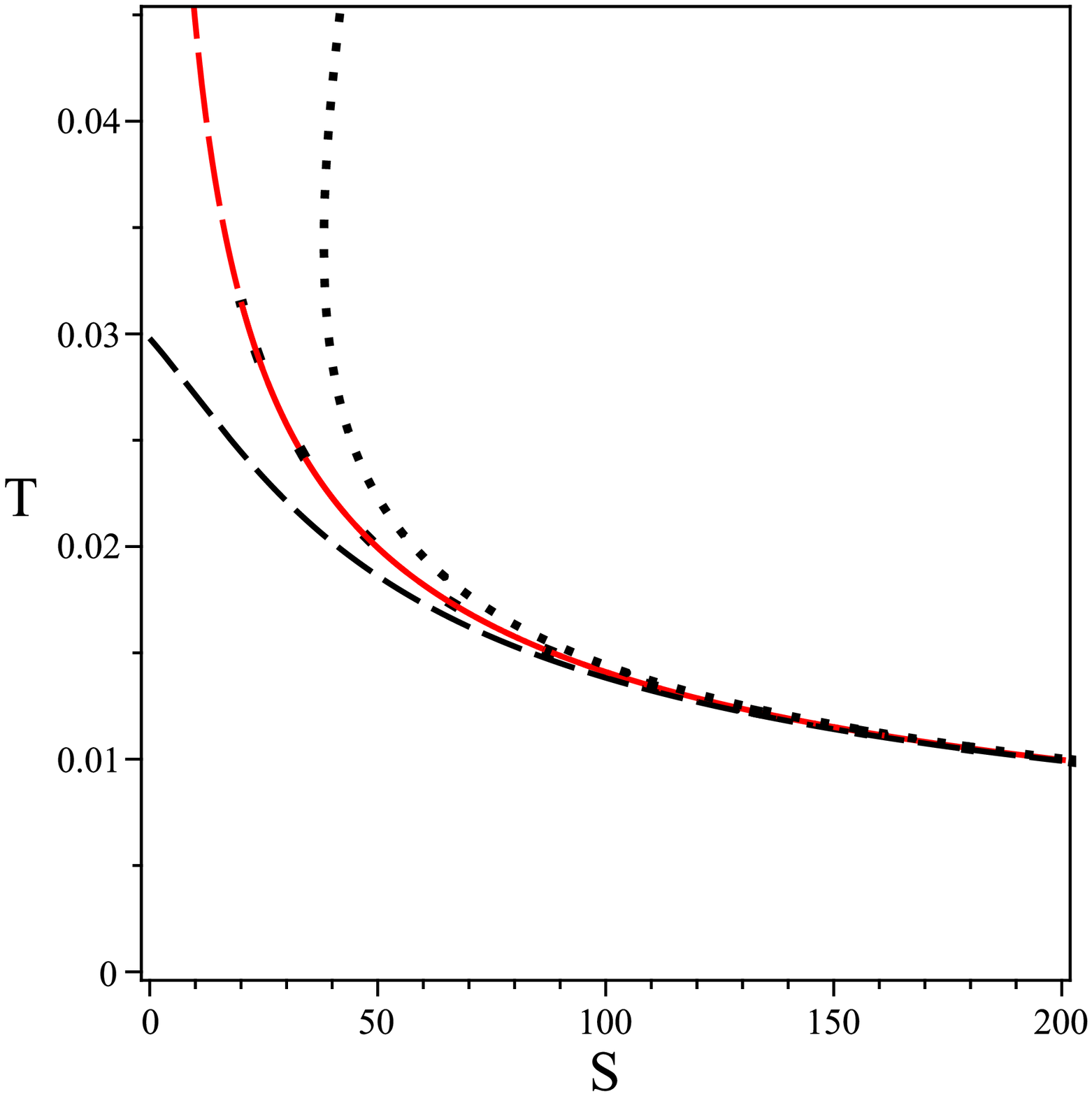}}
\subfigure{\includegraphics[width=0.3\columnwidth]{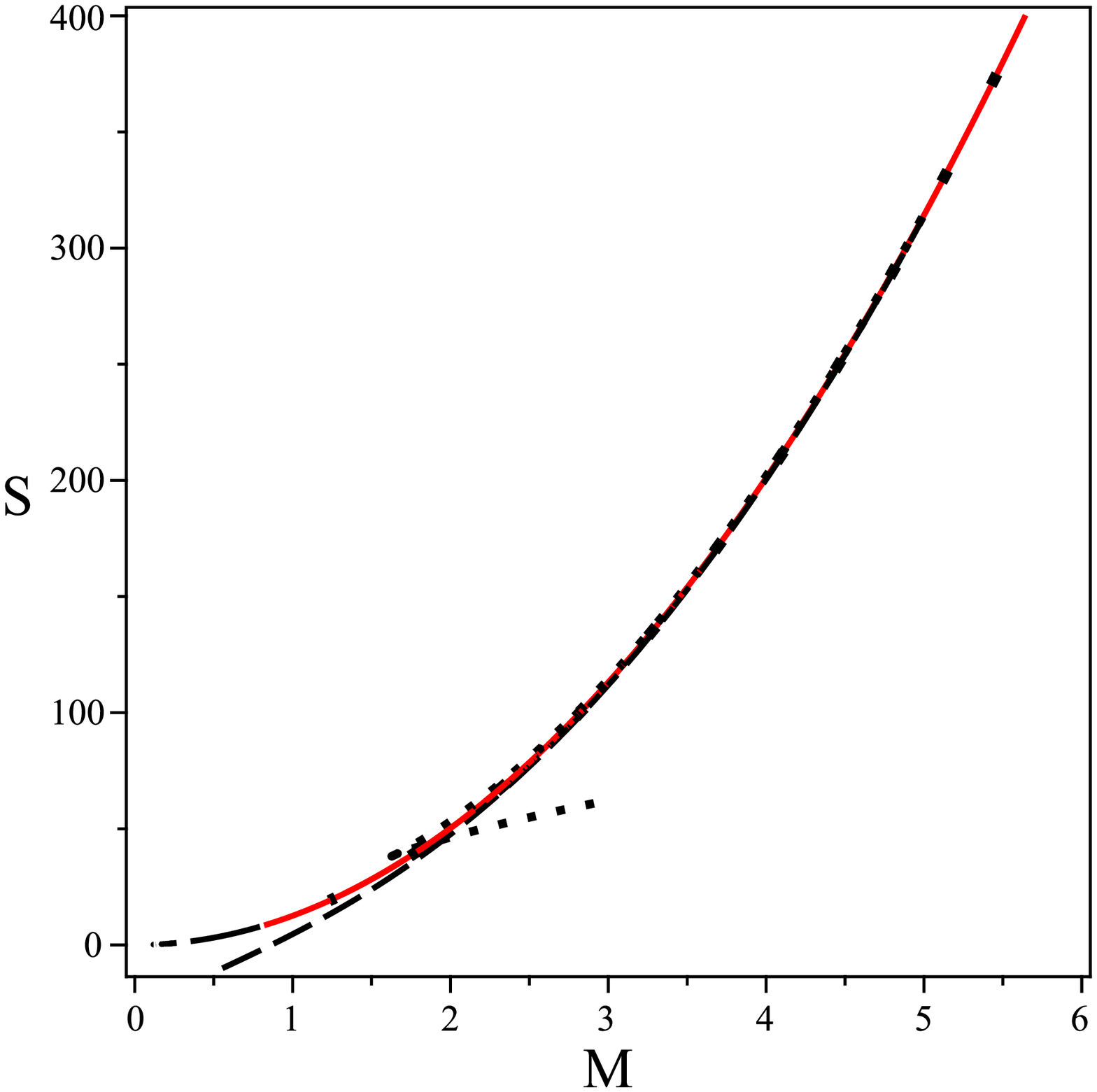}}
\caption{The temperature, entropy and mass as a function of
$r_{+}$ for $c_{1}=-0.25, \alpha=0.5$: solid lines are the results
of our method and dashed lines are the results of the paper
\cite{Bueno:2016lrh}-\cite{Hennigar:2018hza}. For $\alpha=-0.5$: Long
dashed lines are the results of our method and  dotted lines are
for the paper \cite{Bueno:2016lrh}-\cite{Hennigar:2018hza}.}
\label{Ss-Mmplot}
\end{figure}

\begin{figure}[H]\hspace{0.4cm}
\centering
\subfigure{\includegraphics[width=0.45\columnwidth]{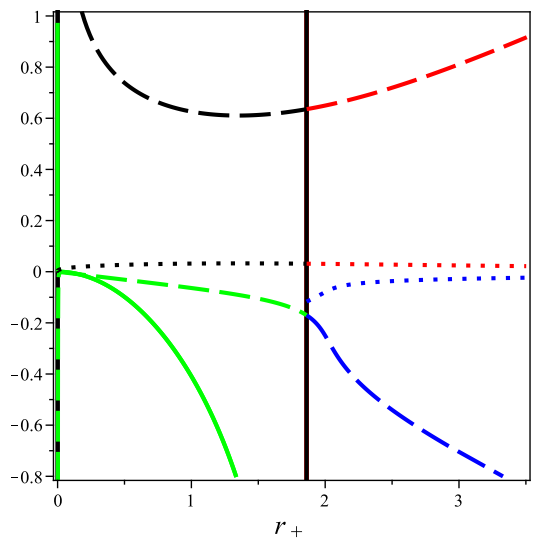}}
\caption{ The temperature (dotted line), heat capacity (solid
line) and free energy (dashed line) as a function of $r_{+}$ for
$c_{1}=-0.25, \alpha=0.5$.} \label{cftrplot}
\end{figure}

\begin{figure}[H]\hspace{0.4cm}
\centering
\subfigure[$r_{+}=6.21,M=3.11$]{\includegraphics[width=0.3\columnwidth]{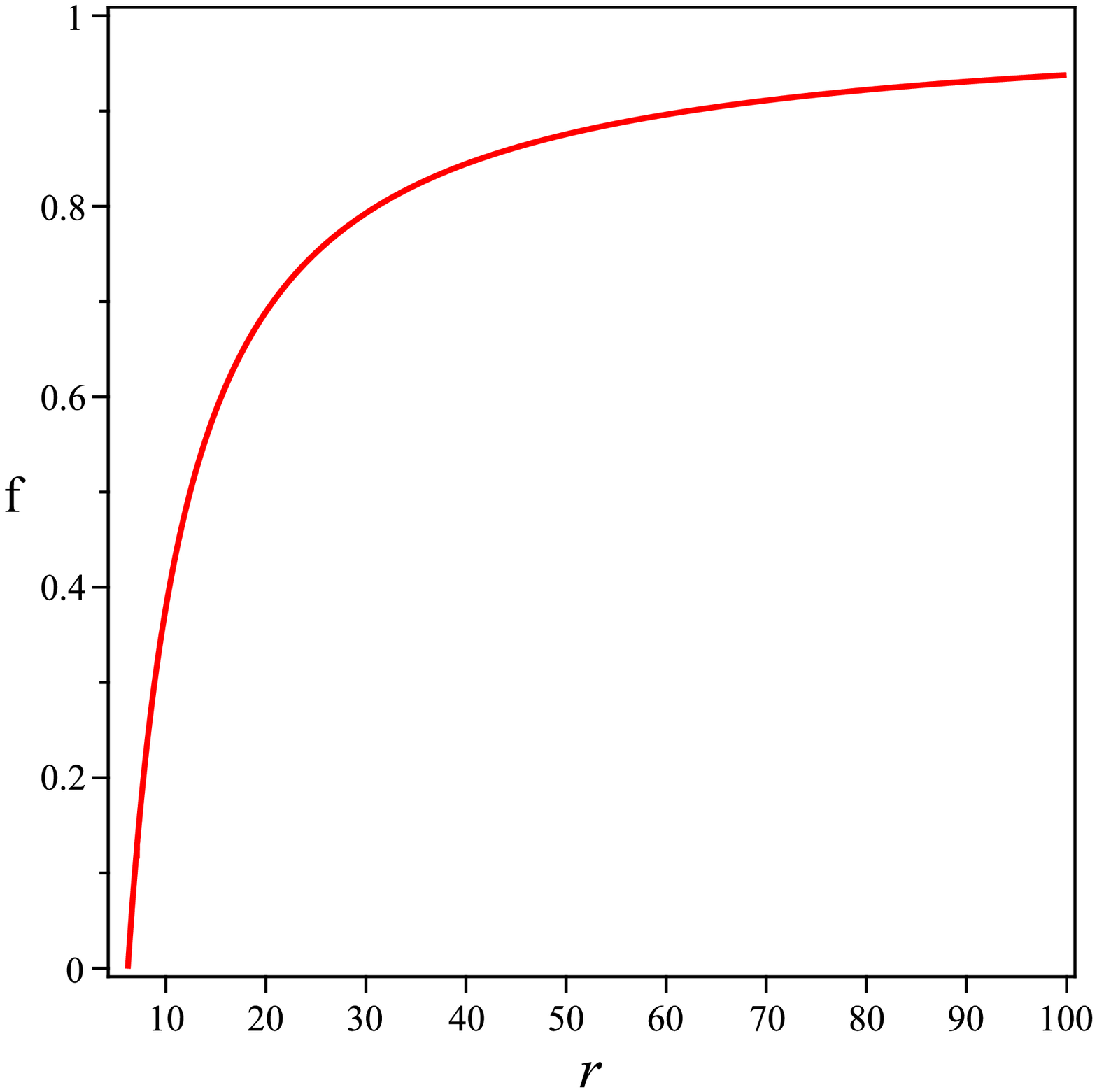}}
\subfigure[$near\; horizon$]{\includegraphics[width=0.3\columnwidth]{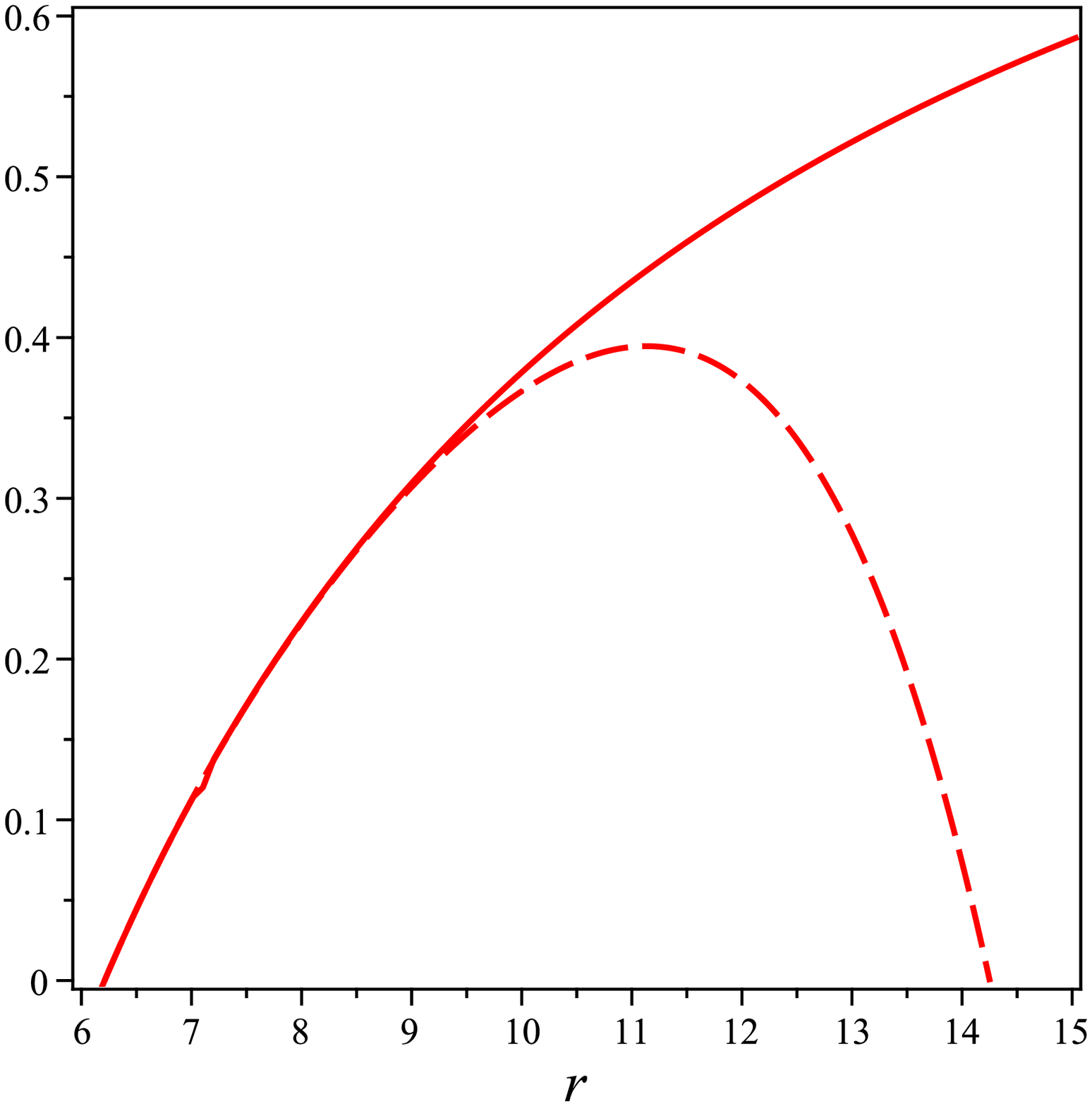}}
\subfigure[$asymptotic$]{\includegraphics[width=0.3\columnwidth]{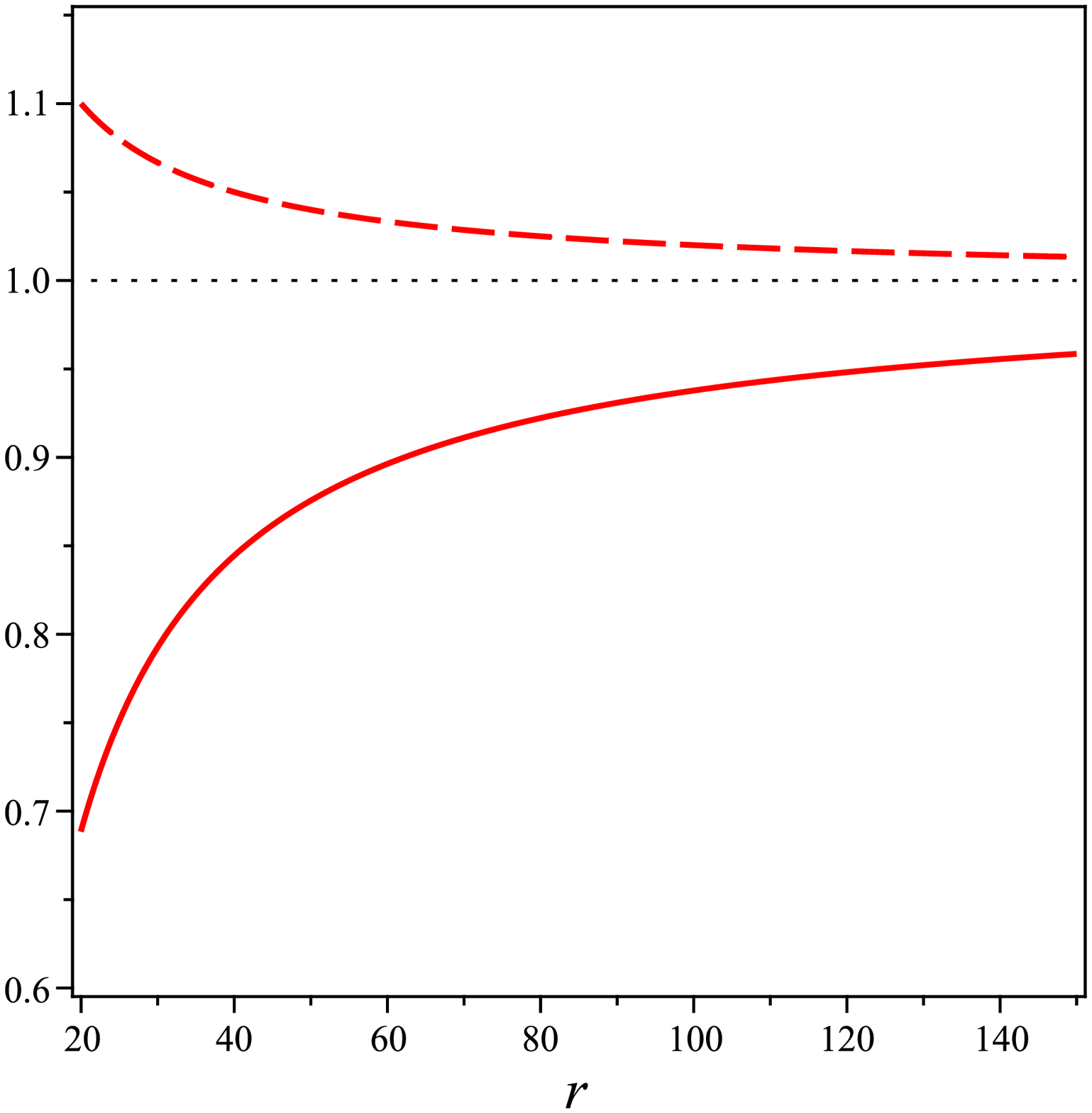}}
\subfigure[$r_{+}=1.0,M=1.28$]{\includegraphics[width=0.3\columnwidth]{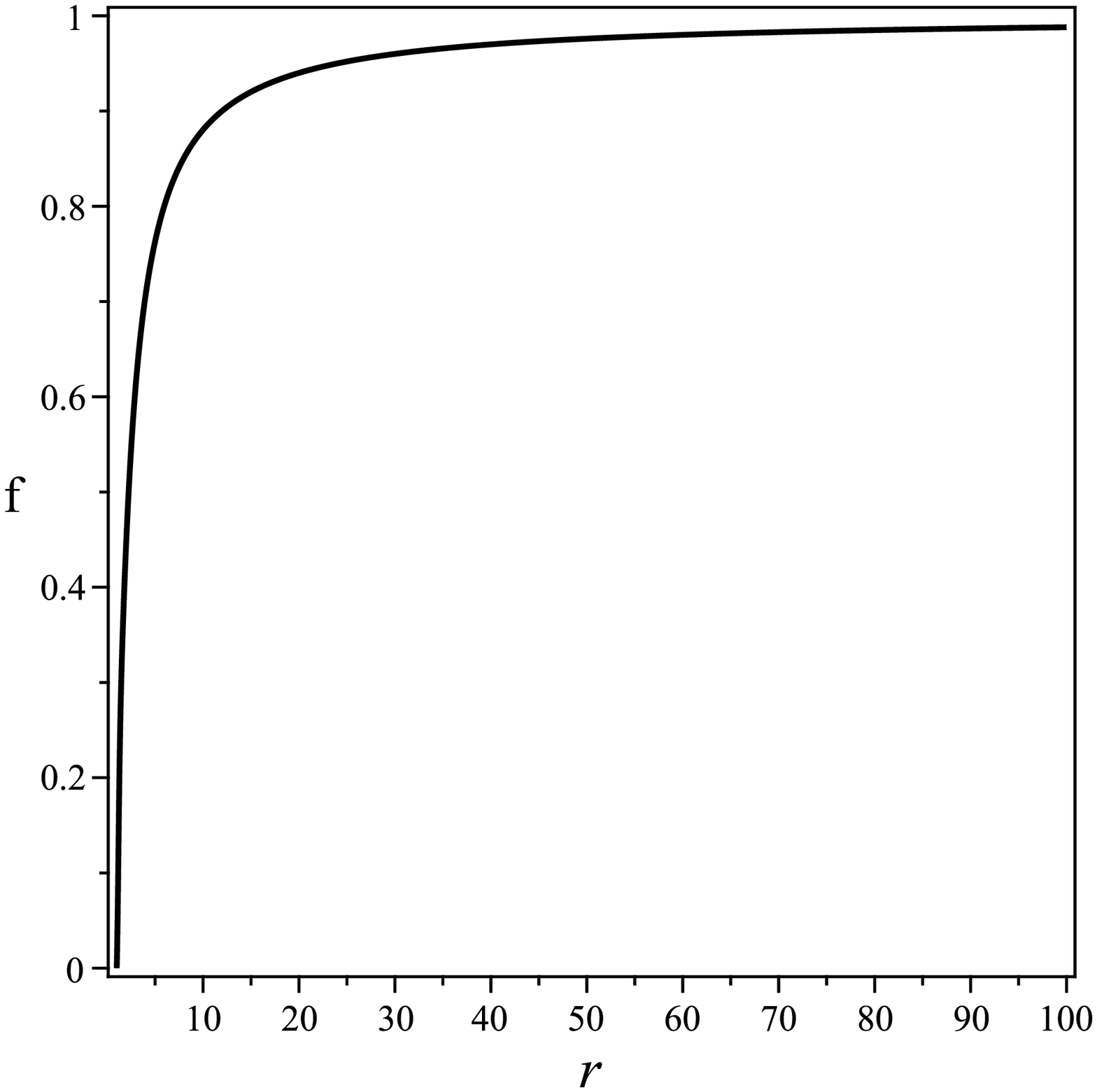}}
\subfigure[$near\; horizon$]{\includegraphics[width=0.3\columnwidth]{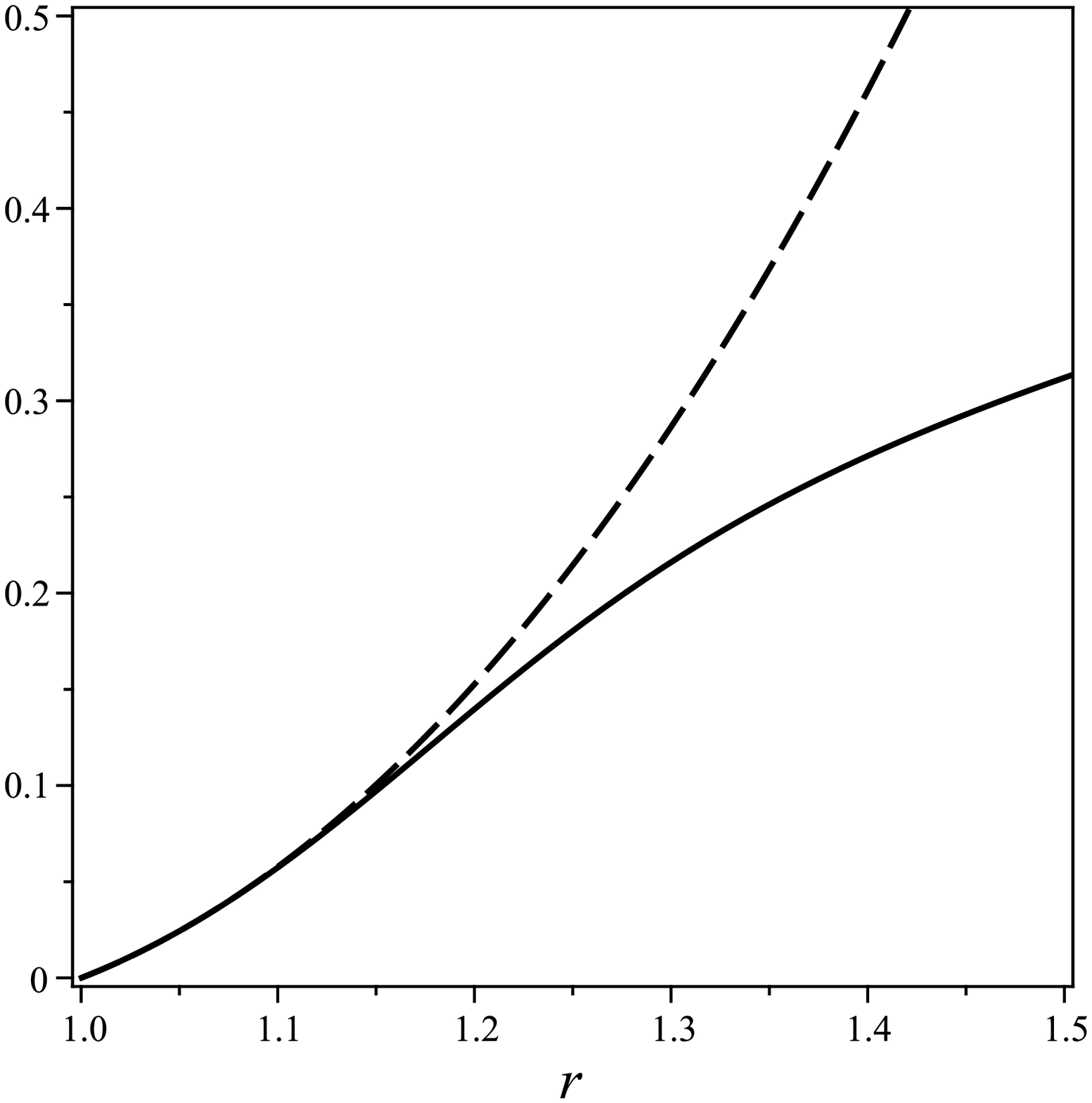}}
\subfigure[$asymptotic$]{\includegraphics[width=0.3\columnwidth]{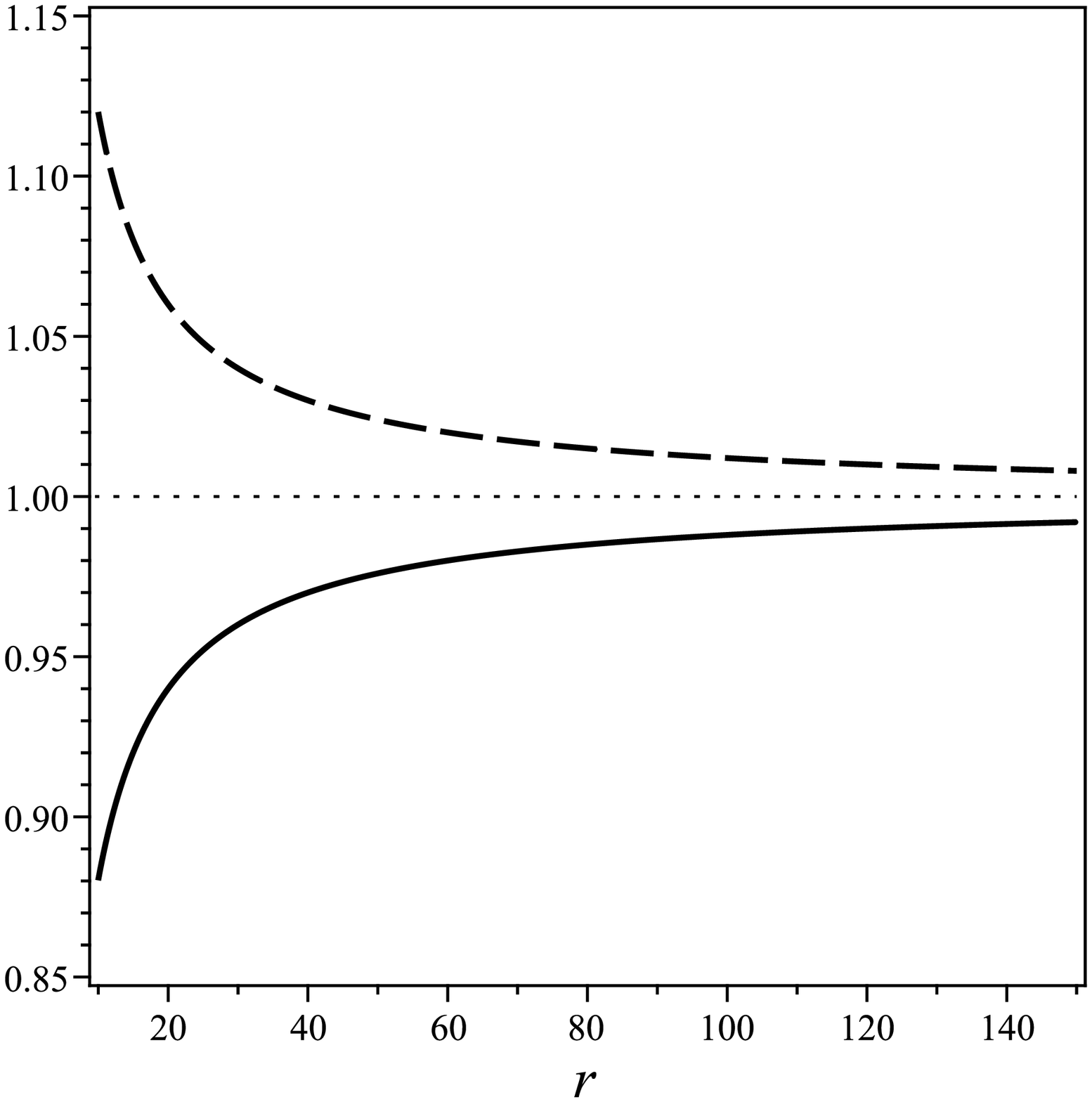}}
\subfigure[$r_{+}=1.0,M=-0.11$]{\includegraphics[width=0.3\columnwidth]{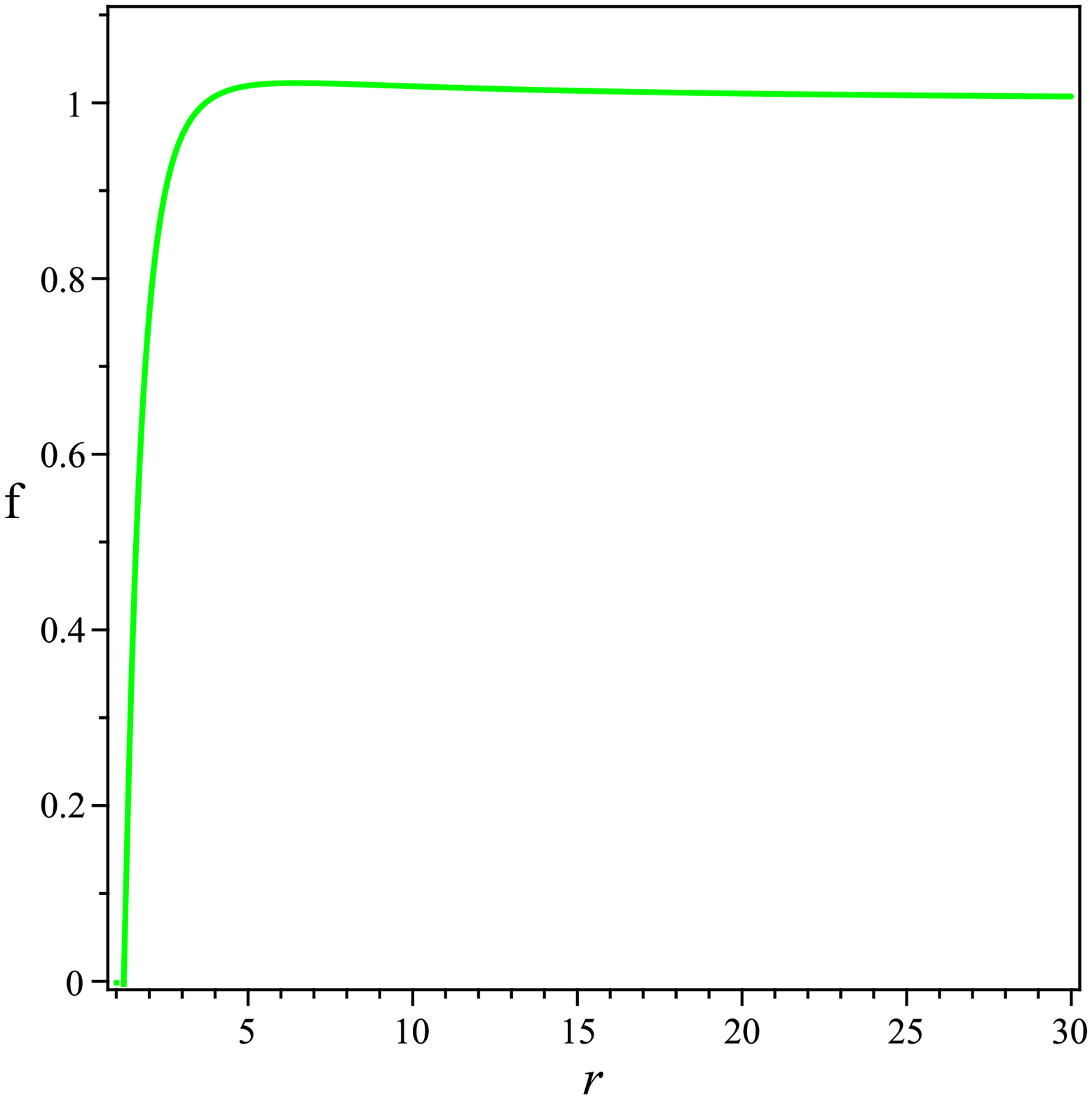}}
\subfigure[$near\; horizon$]{\includegraphics[width=0.3\columnwidth]{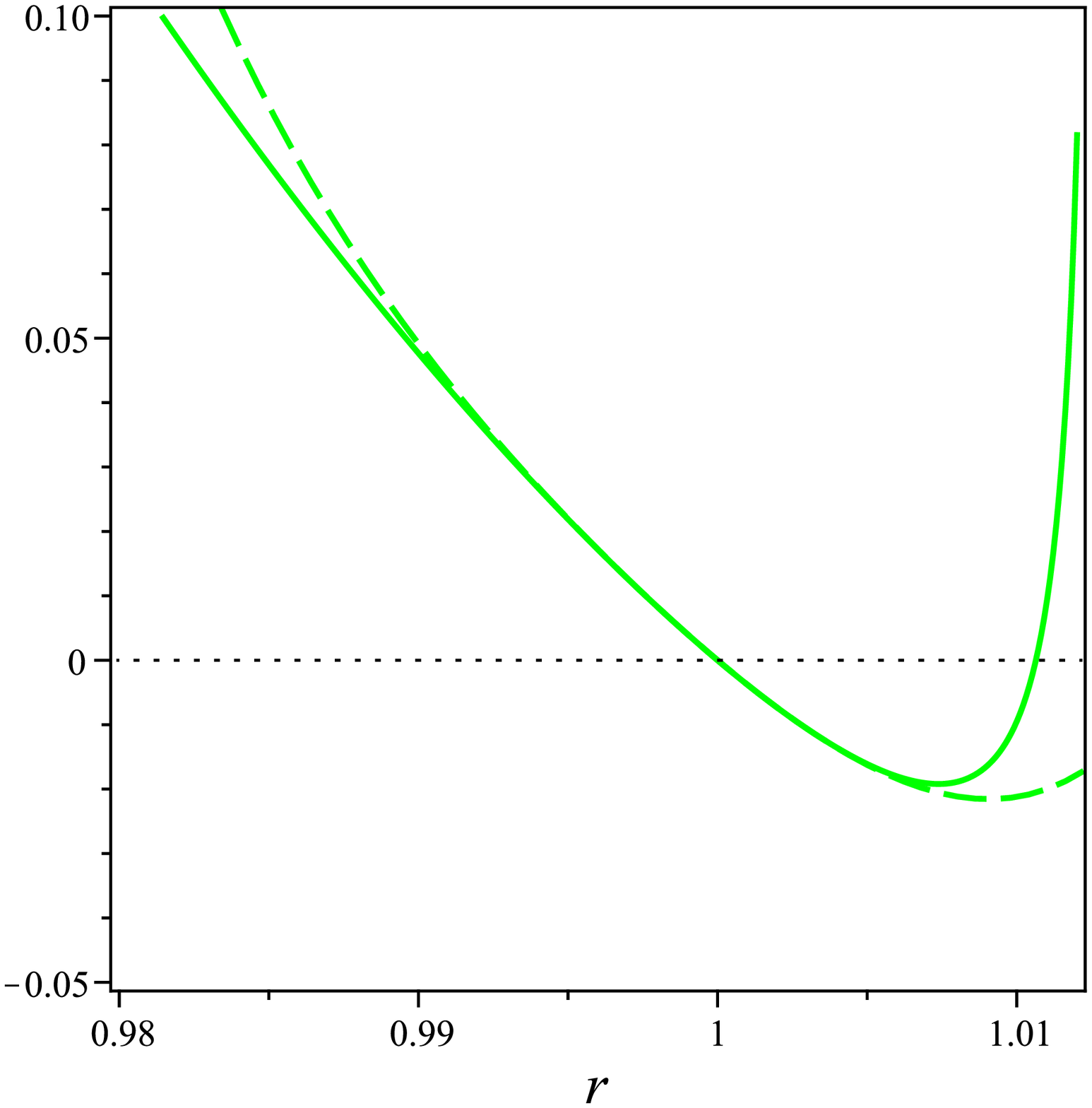}}
\subfigure[$asymptotic$]{\includegraphics[width=0.3\columnwidth]{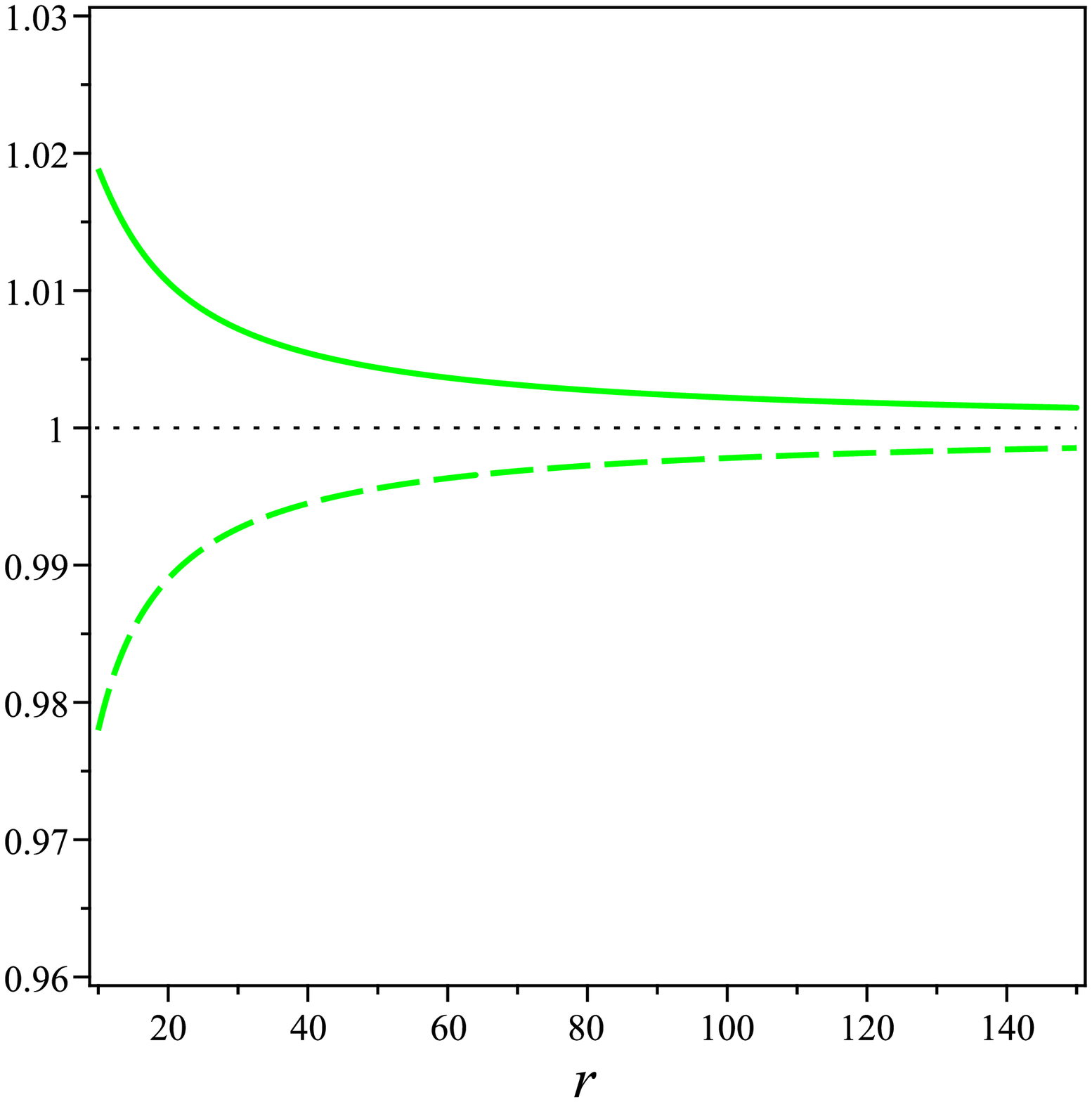}}
\caption{The plots of metric for $c_{1}=-0.25, \alpha=0.5$.}
\label{fplot}
\end{figure}

In panels of Fig. \ref{fplot}, we have depicted the metric
functions. In Figs. \ref{fplot}a and b, the mass of the black hole
is positive and these solutions are similar to Schwarzschild's
black hole. In Figs (\ref{fplot})c and d, the mass of black holes
are negative and their behavior is different from the
Schwarzschild black hole.

\section{New Solutions}\label{sec3}
Here, in order to obtain the new static, spherically symmetric
black hole solutions of theory we consider a line element with
different metric functions. We consider the following static
metric
\begin{equation}\label{metform}
ds^{2}=-h(r)dt^{2}+\dfrac{dr^{2}}{f(r)}+r^2\left(d\theta^{2}+\sin^{2}(\theta)d\phi^{2}\right).
\end{equation}
Unlike the previous section, here we have two metric functions that require two components of field equations to obtain them.
{By inserting the metric into the field equations, the differential equations for $f(r)$ and $h(r)$ become
\begin{align}
8r^4h^4 E^{r}_{r}&=h^3r^4f^{\prime}h^{\prime}+4h^4r^3f^{\prime}+8r^2 f h^{4}+8r^3h^3f h^{\prime}+2r^4f h^{3}h^{\prime\prime}-h^{2}r^{4}fh^{\prime 2}-8r^2 h^{4}-\nonumber\\
&\alpha [-12r^2h^{3}h^{\prime}f^{\prime 3}-48r f h^{3}f^{\prime}h^{\prime\prime}+24rhf^{2}h^{\prime 3}-24rh^{3}h^{\prime}f^{\prime 2}-48r f^{2}h^{2}h^{\prime}h^{\prime\prime}+\nonumber\\
&24rf^{2}h^{2}f^{\prime}h^{\prime 2}+24fr^{2}h^{2}h^{\prime 2}f^{\prime 2}-24rhf^{3}h^{\prime 3}-24r^{2}fh^{3}f^{\prime 2}h^{\prime\prime}-48f^{2}h^{3}h^{\prime}f^{\prime}\nonumber\\
&+48rf^{2}h^{3}f^{\prime}h^{\prime\prime}+48rf^{3}h^{2}h^{\prime}h^{\prime\prime}+12r^{2}f^{3}h^{\prime 4}+48fh^{3}h^{\prime}f^{\prime}+24r^{2}f^{2}h^{2}f^{\prime}h^{\prime}h^{\prime\prime}\nonumber\\
&+48rfh^{3}h^{\prime}f^{\prime 2}-24r^{2}hf^{2}f^{\prime}h^{\prime 3}-24r^{2}f^{3}hh^{\prime 2}h^{\prime\prime}],
\end{align}
and
\begin{align}
-4r^{4}h^{4}E^{a}_{a}&=-3r^{4}h^{3}f^{\prime}h^{\prime}-6r^{4}h^{3}fh^{\prime\prime}-
12r^{3}h^{4}f^{\prime}-12r^{2}fh^{4}-12fr^{3}h^{3}h^{\prime}+3r^{4}fh^{2}h^{\prime 2}+12r^{2}h^{4}\nonumber\\
&+3\alpha[-16r^{2}fh^{3}f^{\prime 2}h^{\prime\prime}+32rf^{2}h^{3}f^{\prime}h^{\prime\prime}+32rfh^{3}h^{\prime}f^{\prime 2}-16r^{2}hf^{2}f^{\prime}h^{\prime 3}-16r^{2}hf^{3}h^{\prime\prime}h^{\prime 2}\nonumber\\
&+16rf^2h^2f^{\prime}h^{\prime 2}+16r^2f^2h^2f^{\prime}h^{\prime}h^{\prime\prime}+16r^2fh^{2}f^{\prime 2}h^{\prime 2}-8r^2h^{3}h^{\prime}f^{\prime 3}+8r^2f^{3}h^{\prime 4}-16rhf^{3}h^{\prime 3}\nonumber\\
&+32rh^{2}f^{3}h^{\prime}h^{\prime\prime}-32f^{2}h^{3}h^{\prime}f^{\prime}-32rf^{2}h^{2}h^{\prime}h^{\prime\prime}-
32rfh^{3}f^{\prime}h^{\prime\prime}+16rhf^{2}h^{\prime 3}-16rh^{3}f^{\prime 2}h^{\prime}\nonumber\\
&+32fh^{3}h^{\prime}f^{\prime}].
\end{align}
}
Similar to the previous section, expanding the function $h(r)$ and $f(r)$ around the event horizon $r_+$
\begin{align}\label{eq72}
f(r)&= f_{1}(r-r_{+})+f_{2}(r-r_{+})^{2}+f_{3}(r-r_{+})^{3}+...\\
h(r)&=h_{1}(r-r_{+})+h_{2}(r-r_{+})^{2}+h_{3}(r-r_{+})^{3}+...
\end{align}
and then inserting these expressions into the field equations, we
find
\begin{align}\label{eq92}
f_{2}={\dfrac{h_{1}r_{+}^2-h_{1}f_{1}r_{+}^3+6\alpha f_{1}^3h_{2}r_{+}^2+6\alpha h_{1}f_{1}^2+12\alpha r_{+}h_{2}f_{1}^2}{6\alpha r_{+}f_{1}h_{1}(2+r_{+}f_{1})},}
\end{align}
where $r_+$, $f_1$, $h_{1}$ and $h_{2}$ are undetermined constants
of integration. The other near horizon constants provided in the
Appendixes \ref{appb}.
{ In the large $r$ limit, we linearize the field equations near the Schwarzschild background
\begin{align}
f(r)&=1-\dfrac{2M}{r}+\epsilon F(r),\\
h(r)&=1-\dfrac{2M}{r}+\epsilon H(r),
\end{align}
where $F(r)$ and $H(r)$ are determined by the field equations, and we linearize the differential equations by keeping terms only to order $\epsilon$. The resulting differential equations for $F(r)$ and $H(r)$ take the form
\begin{align}
H^{\prime\prime}+\gamma(r)H^{\prime}+\eta(r)F^{\prime}+\omega(r)F(r)+\Xi(r)H(r)+g(r)&=0,\label{eqasymp45}\\
H^{\prime\prime}+\bar{\gamma}(r)H^{\prime}+\bar{\eta}(r)F^{\prime}+\bar{\omega}(r)F(r)+\bar{\Xi}(r)H(r)+\bar{g}(r)&=0,\label{eqasymp46}
\end{align}
where the functions shown in the above equations are given in the appendix \ref{app3}.
In the large $r$ limit, the homogenous equations read
\begin{align}
H^{\prime\prime}+\dfrac{4}{r}H^{\prime}+\dfrac{2}{r}F^{\prime}+\dfrac{4F}{r^2}-\dfrac{4MH}{r^3}&=0,\label{eqq47}\\
H^{\prime\prime}+\dfrac{2}{r}H^{\prime}+\dfrac{2}{r}F^{\prime}+\dfrac{2F}{r^2}+\dfrac{2M^{2}H}{r^4}&=0.\label{eqq48}
\end{align}
Equations \eqref{eqq47} and \eqref{eqq48} can be solved exactly to obtain
\begin{align}
F(r)&=\dfrac{M}{\sqrt{\pi}r^2}\left[2c_{2}^{\prime}re^{-\dfrac{(M+2r)^2}{r^2}}+\sqrt{\pi}(M+2r)\left(c_{1}^{\prime}+c_{2}^{\prime}{\it erf}\left(2+\dfrac{M}{r}\right)\right)\right],\\
H(r)&=c_{1}+c_{2}{\it erf}\left(2+\dfrac{M}{r}\right).
\end{align}
In the large $r$, $F(r)$ and $H(r)$ become 
\begin{align}
F(r)\approx &\dfrac{2M(\sqrt{\pi}c_{1}^{\prime}+\sqrt{\pi}c_{2}^{\prime}{\it erf}(2)+c_{2}^{\prime}e^{-4})}{\sqrt{\pi}r}+\dfrac{M^{2}(\sqrt{\pi}c_{1}^{\prime}+\sqrt{\pi}c_{2}^{\prime}{\it erf}(2)-c_{2}^{\prime}e^{-4})}{\sqrt{\pi}r^{2}}+\dfrac{8c_{2}^{\prime}e^{-4}M^{3}}{\sqrt{\pi}r^{3}}\nonumber\\
&-\dfrac{8c_{2}^{\prime}e^{-4}M^{4}}{\sqrt{\pi}r^{4}}+\mathcal{O}(r^{-5}),\\
H(r)\approx &c_{1}+c_{2}{\it erf}(2)+\dfrac{2c_{2}e^{-4}M}{\sqrt{\pi}r}-\dfrac{4c_{2}e^{-4}M^{2}}{\sqrt{\pi}r^{2}}+\dfrac{14c_{2}e^{-4}M^{3}}{3\sqrt{\pi}r^{3}}-\dfrac{10c_{2}e^{-4}M^{4}}{3\sqrt{\pi}r^{4}}+\mathcal{O}(r^{-5}).
\end{align}
As we know, for $\alpha\to 0$, the metric is expected to return to the Schwarzschild metric. To fulfill this desire, we must have: $c_{1}=c_{1}^{\prime}=c_{2}=c_{2}^{\prime}=0$. To obtain  the particular solution, we consider the following expansions
\begin{align}\label{eqqqhfo}
 f_{p}(r)=&\sum_{n=2}\dfrac{F_{n}}{r^{n}}=\dfrac{F_{2}}{r^2}+\dfrac{F_{3}}{r^3}...\\
 h_{p}(r)=&\sum_{n=2}\dfrac{H_{n}}{r^{n}}=\dfrac{H_{2}}{r^2}+\dfrac{H_{3}}{r^3}...
\end{align}
Inserting the above expansions into the field equations and solving order by order, one can get
\begin{align}
f_{p}(r)&=-\dfrac{1120\alpha M^{3}}{3r^{7}}-\dfrac{288\alpha M^{4}}{r^{8}}-\dfrac{756\alpha M^{5}}{r^{9}}-\dfrac{2016\alpha M^{6}}{r^{10}}-\dfrac{27216\alpha M^{7}}{5r^{11}}+\mathcal{O}\left(\dfrac{1}{r^{12}}\right)\\
h_{p}(r)&=-\dfrac{256\alpha M^{3}}{3r^{7}}-\dfrac{108\alpha M^{4}}{r^{8}}-\dfrac{252\alpha M^{5}}{r^{9}}-\dfrac{3024\alpha M^{6}}{5r^{10}}-\dfrac{81648\alpha M^{7}}{55r^{11}}+\mathcal{O}\left(\dfrac{1}{r^{12}}\right).
\end{align}
In large $r$, the solutions are:
\begin{equation}\label{eqqasymp}
f(r)\approx 1-\dfrac{2M}{r}+f_{p},\;\;\;\;\;h(r)\approx 1-\dfrac{2M}{r}+h_{p}.
\end{equation}
}
We wish to obtain an approximate analytic solution that is valid
near the horizon and at large $r$.  To this end we employ a
continued fraction expansion, and write
\begin{equation}\label{eqq112}
h(r)=xA(x),\hspace{0.5cm}\dfrac{h(r)}{f(r)}=B^{2}(x),
\end{equation}
with
\begin{align}
A(x) &=1-\epsilon(1-x)+(a_{0}-\epsilon)(1-x)^{2}+\tilde{A}(x)(1-x)^{3}
\label{Ax}
\\
B(x) &=1+b_{0}(1-x)+\tilde{B}(x)(1-x)^{2}
\label{Bx}
\end{align}
where
\begin{equation}
x = 1- \frac{r_+}{r} \qquad
\tilde{A}(x)=\dfrac{a_{1}}{1+\dfrac{a_{2}x}{1+\dfrac{a_{3}x}{1+\dfrac{a_{4}x}{1+...}}}}
\qquad
\tilde{B}(x)=\dfrac{b_{1}}{1+\dfrac{b_{2}x}{1+\dfrac{b_{3}x}{1+\dfrac{b_{4}x}{1+...}}}}
\label{cfrac}
\end{equation}
where we truncate the continued fraction at order $4$.
By expanding (\ref{eqq112}) near the horizon ($ x\to 0 $) and
the asymptotic  region ($ x\to 1 $)  we obtain
\be
\epsilon=-\dfrac{H_{1}}{r_{+}}-1,  \qquad b_{0}=\dfrac{F1-H1}{2r_{+}},  \qquad a_{0}=\dfrac{H_{2}}{r_{+}^{2}}
\ee
for the lowest order expansion coefficients, with the remaining
$a_i$ and $b_i$ given in terms of $(r_+, h_1, f_1, h_{2})$; we provide these expressions in the  Appendix.
 For a static space time we have a timelike Killing vector
$ \xi=\partial_{t} $ everywhere outside the horizon and so  we obtain
\begin{align}\label{eq400}
T =&\dfrac{1}{4\pi}\left. \sqrt{\dfrac{f(r)}{h(r)}}h^{'}(r)\right\vert_{r_{+}}
= \dfrac{\sqrt{f_{1}h_{1}}}{4\pi} = {\dfrac {(1-2\epsilon+a_{1}+a_{0})}{{4\pi r_{+}} \left( 1+{ b_{1}} \right) }}.
\end{align}
We compute the entropy by using of the first two term in continued
fraction expansion as follows \cite{Wald1,Wald2}
\begin{align}
S=&-2\pi\int_{Horizon}d^{2}x\sqrt{\eta}\dfrac{\delta L}{\delta R_{a b c d}}\epsilon_{a b}\epsilon_{c d}=
\pi r_{+}^{2}\left[1+6\alpha\left(\dfrac{2h_{1}}{r_{+}^3}-\dfrac{f_{1}h_{1}}{r_{+}^2}+\dfrac{2h(r_{+})f_{1}}{f(r_{+})r_{+}^3}+\dfrac{h(r_{+})f_{1}^2}{f(r_{+})r_{+}^2}+\dfrac{f(r_{+})h_{1}^2}{h(r_{+})r_{+}^2}\right)\right]\nonumber\\
&=\pi r_{+}^2\left[1+6\alpha\left(\dfrac{4h_{1}}{r_{+}^3}+\dfrac{f_{1}h_{1}}{r_{+}^2}\right)\right].
\label{eq12}
\end{align}
Then the mass from \eqref{eqfirstlaw}, becomes
\begin{equation}
M=\dfrac{\sqrt{f_{1}h_{1}}r_{+}^2}{2}\left[1+6\alpha \left(\dfrac{4h_{1}}{r_{+}^3}+\dfrac{f_{1}h_{1}}{r_{+}^2}\right)\right].
\end{equation}
To obtain the new black hole solutions we consider the different
relations between $f_{1}$ and $h_{1}$ \cite{Bonanno:2019rsq}. In the case of
$f_{1}=h_{1}$, we obtain the results of the first section.

\subsection{The case $h_{1}=f_{1}^{3}$}
\begin{figure}[H]\hspace{0.4cm}
\centering
\subfigure{\includegraphics[width=0.3\columnwidth]{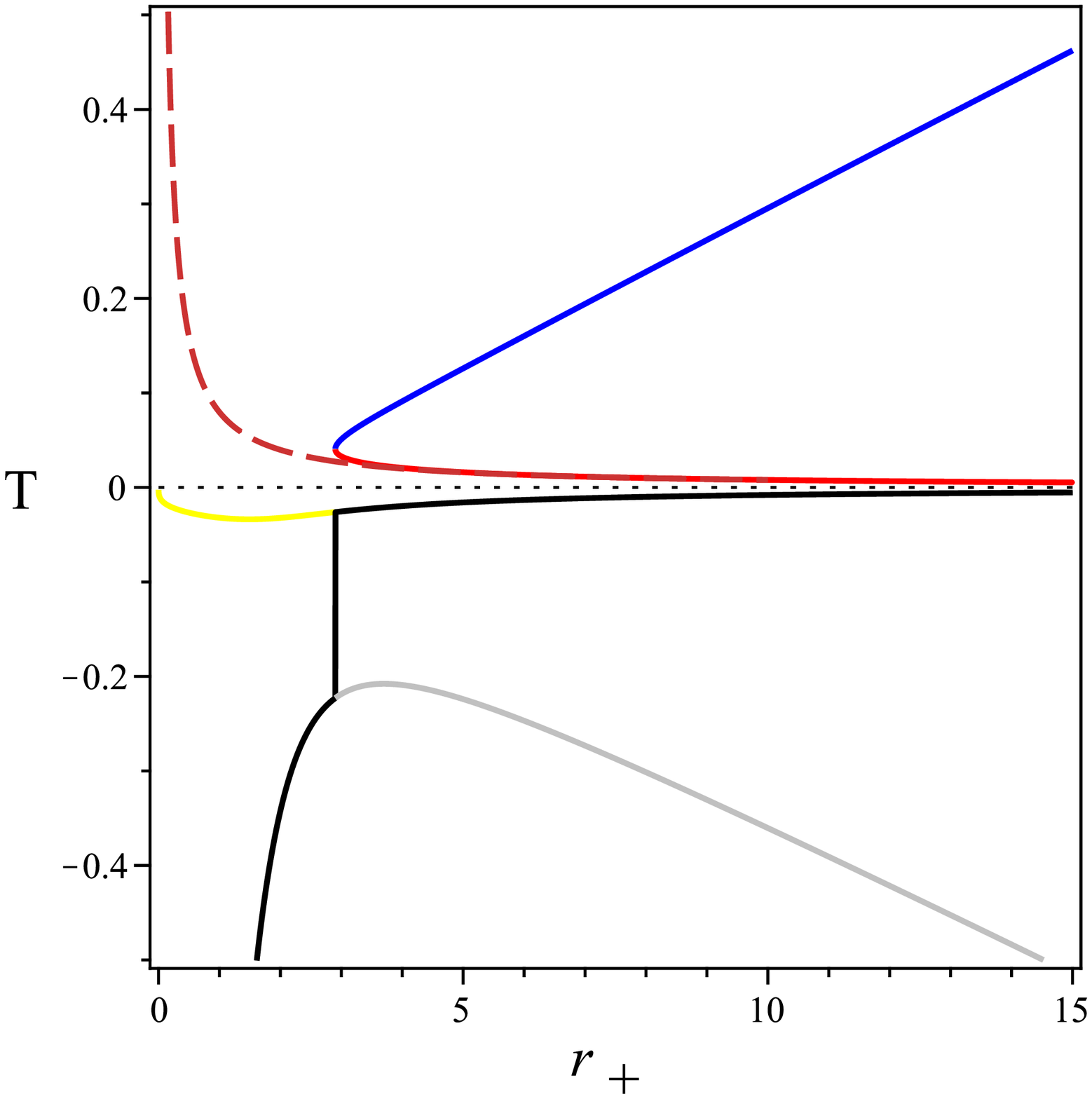}}
\subfigure{\includegraphics[width=0.3\columnwidth]{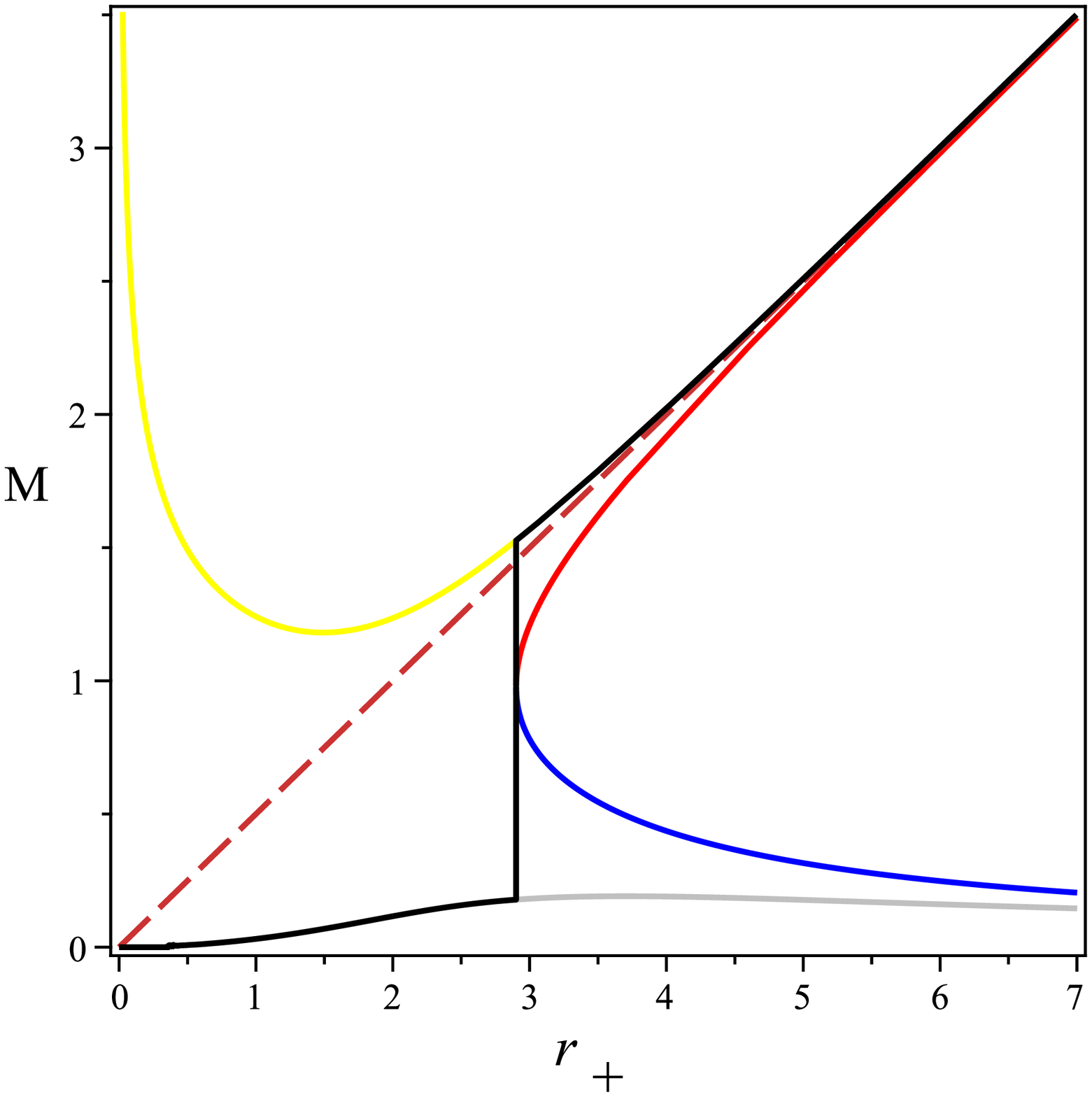}}
\subfigure{\includegraphics[width=0.3\columnwidth]{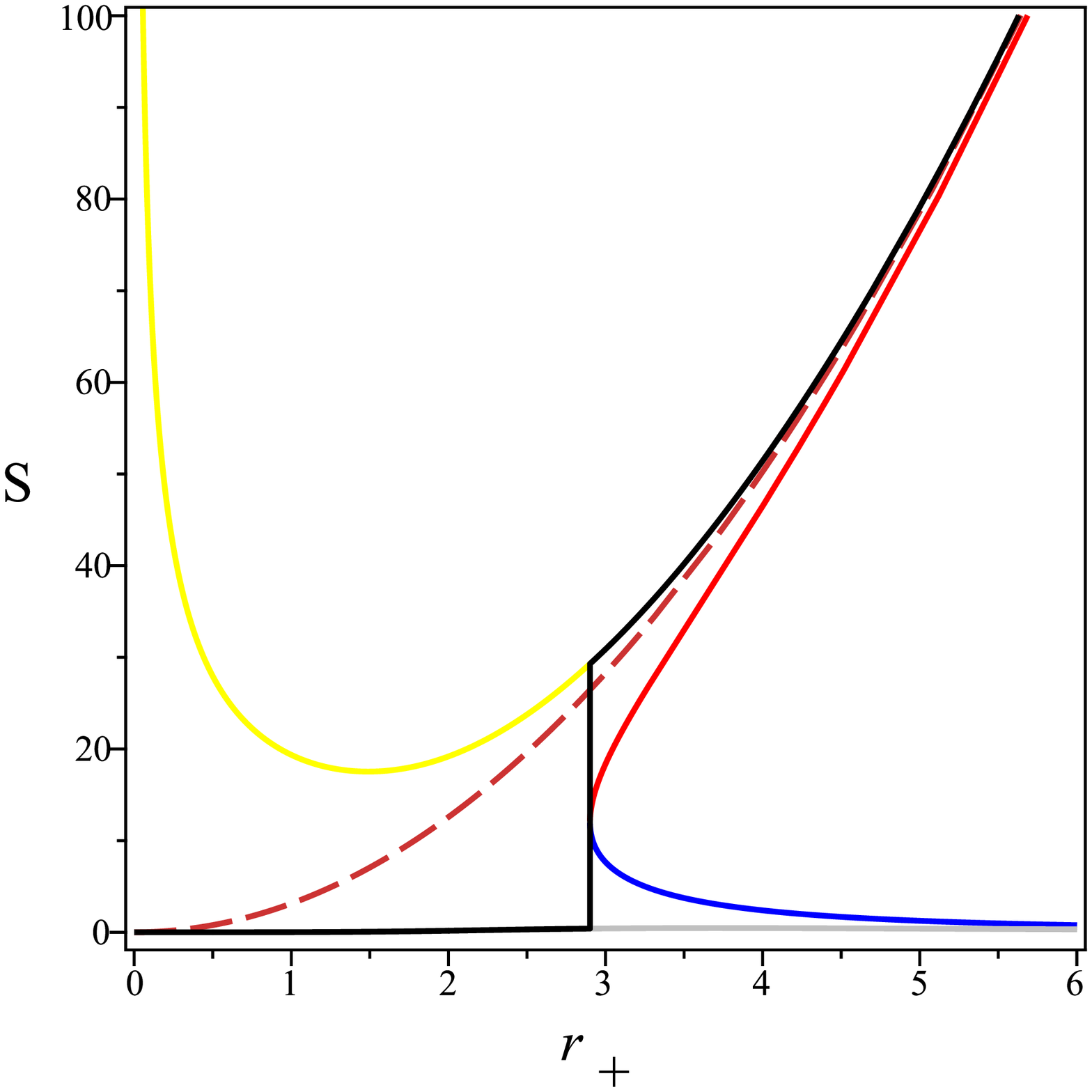}}
\caption{The temperature, entropy and mass as a function of
$r_{+}$ for $c_{1}=\alpha=-0.5$. The orange dashed lines related
to the Schwarzschild's black hole. } \label{TMSplott}
\end{figure}

\begin{figure}[H]\hspace{0.4cm}
\centering
\subfigure{\includegraphics[width=0.4\columnwidth]{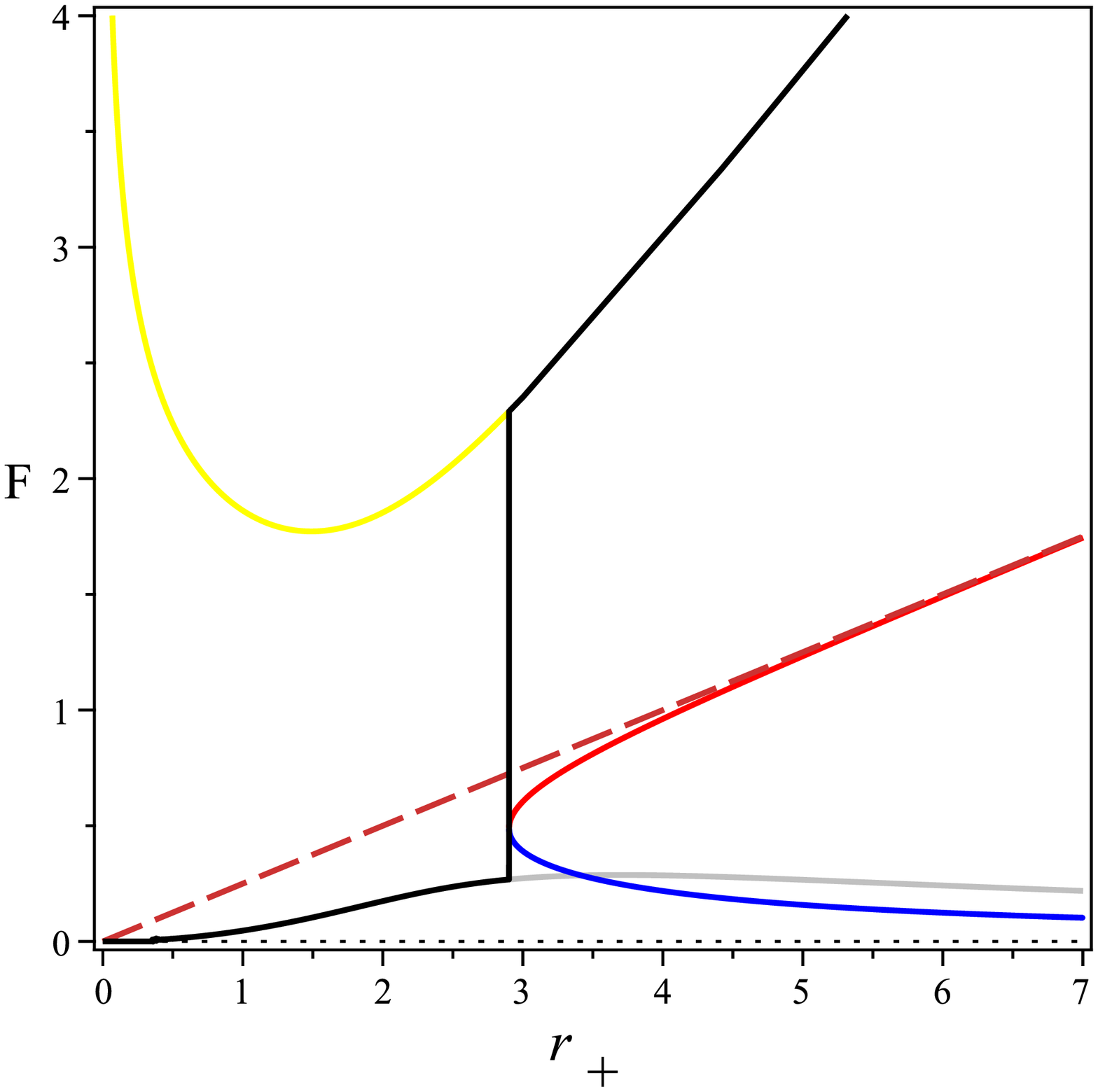}}
\subfigure{\includegraphics[width=0.4\columnwidth]{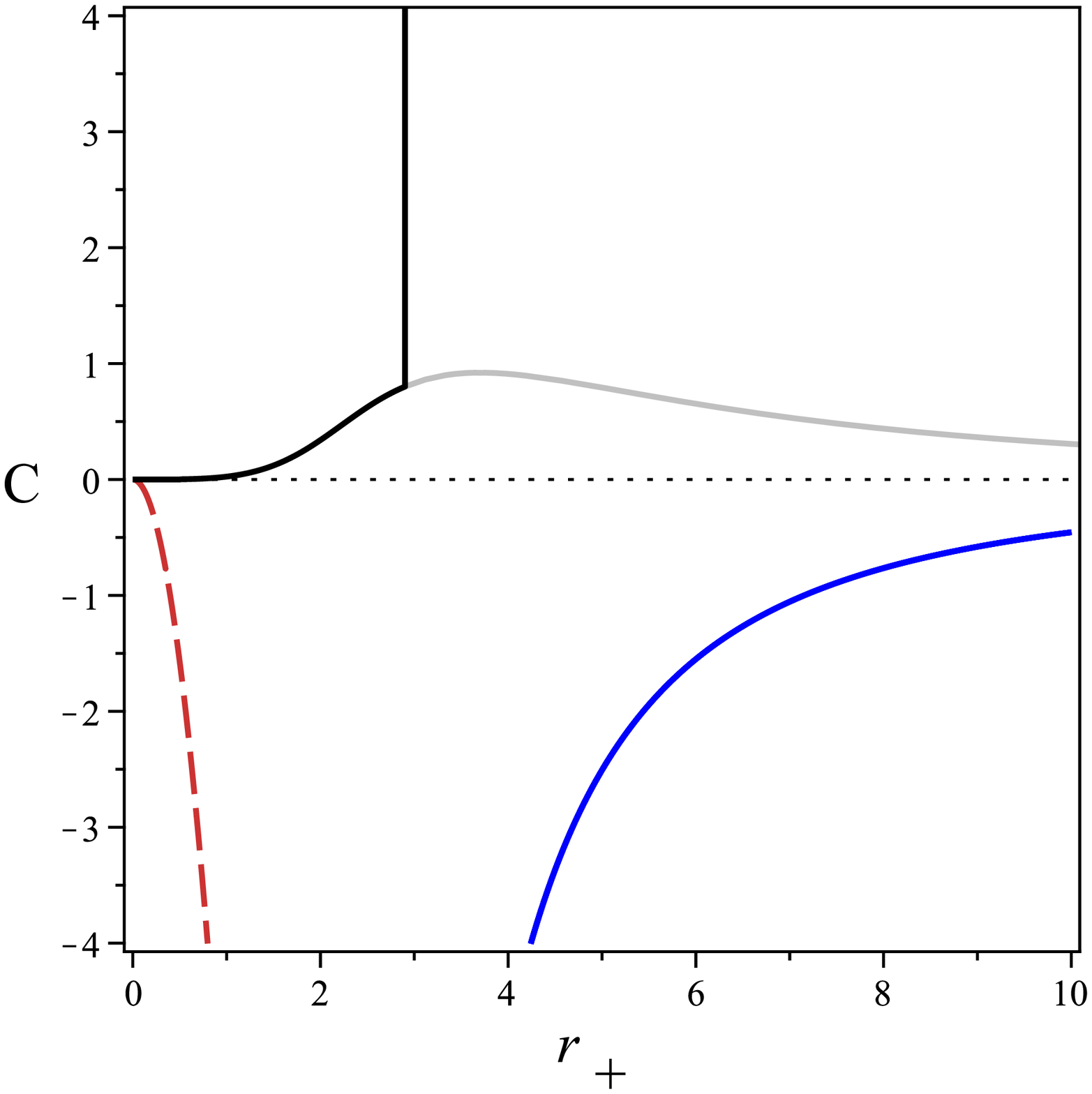}}
\caption{The free energy and heat capacity as a function of
$r_{+}$ for $c_{1}=\alpha=-0.5$. The orange dashed lines related
to the Schwarzschild's black hole.} \label{CFplott}
\end{figure}

\begin{figure}[H]\hspace{0.4cm}
\centering
\subfigure{\includegraphics[width=0.5\columnwidth]{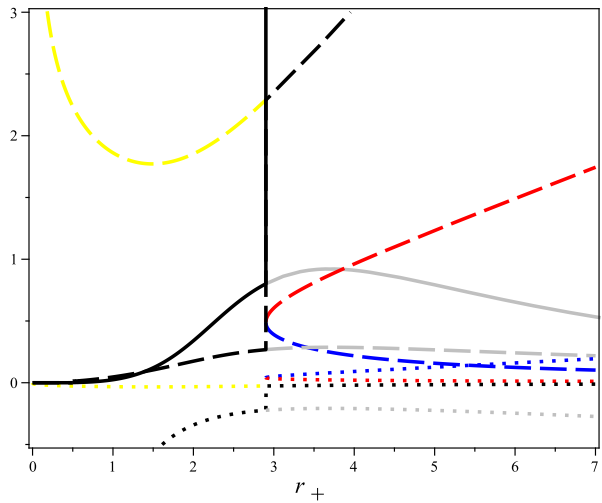}}
\caption{ The temperature (dotted line), heat capacity (solid
line) and free energy (dashed line) as a function of $r_{+}$ for
$c_{1}=-0.25, \alpha=-0.5$.} \label{cftrplot2}
\end{figure}

We now consider following relation between $f_1$ and $h_1$ as
\begin{equation}
h_{1}(r_{+})=f_{1}^3(r_{+}).
\end{equation}
From equations \eqref{eqfirstlaw} and \eqref{eq26}, we have
\begin{align}
&M=\dfrac{f_{1}^2}{2r_{+}}(r_{+}^3+12\alpha r_{+}f_{1}^{4}(r_{+})+48\alpha f_{1}^3(r_{+})),\\
&2r_{+}^{4}f_{1}^{\prime}(r_{+})+48\alpha r_{+}^2f_{1}^{4}(r_{+})f_{1}^{\prime}(r_{+})+168\alpha r_{+}f_{1}^{3}(r_{+})f^{\prime}_{1}(r_{+})+f_{1}(r_{+})r_{+}^3-24\alpha f_{1}^{4}(r_{+})=0.\label{eqq45q}
\end{align}
Solving (\ref{eqq45q}), one can achieve following equation as
\begin{equation}\label{eqq46q}
\dfrac{1}{2}r_{+}^2f_{1}^4+\dfrac{24\alpha f_{1}^{7}}{r_{+}}+6\alpha f_{1}^{8}+c_{1}=0
\end{equation}
which leads to a maximum of eight solutions for $f_{1}(r_{+})$
depending on the value of parameters. Inserting the solutions of
\eqref{eqq46q} into the thermodynamical quantities, one can obtain
the analytical solutions for them, which we have plotted in Figs.
\ref{TMSplott} and \ref{CFplott}.  In these figures, the dashed
orange lines are the behavior of the thermodynamical quantities of
Schwarzschild's black hole. In Fig. (\ref{TMSplott}), the blue and
red solid lines are the physical branches, because the
temperatures and the mass of these branches are positive. Also,
the red branch is globally and locally unstable and the blue
branch is globally stable while locally unstable (Fig.
(\ref{CFplott})). From the Fig. (\ref{cftrplot2}), there is a
divergence in the heat capacity, but this divergence does not
coincide with the extremum points of temperature and free energy.
So, the phase transition does not occur.

\begin{figure}[H]\hspace{0.4cm}
\centering \subfigure[$h_{2}=1,
r_{+}=6$]{\includegraphics[width=0.3\columnwidth]{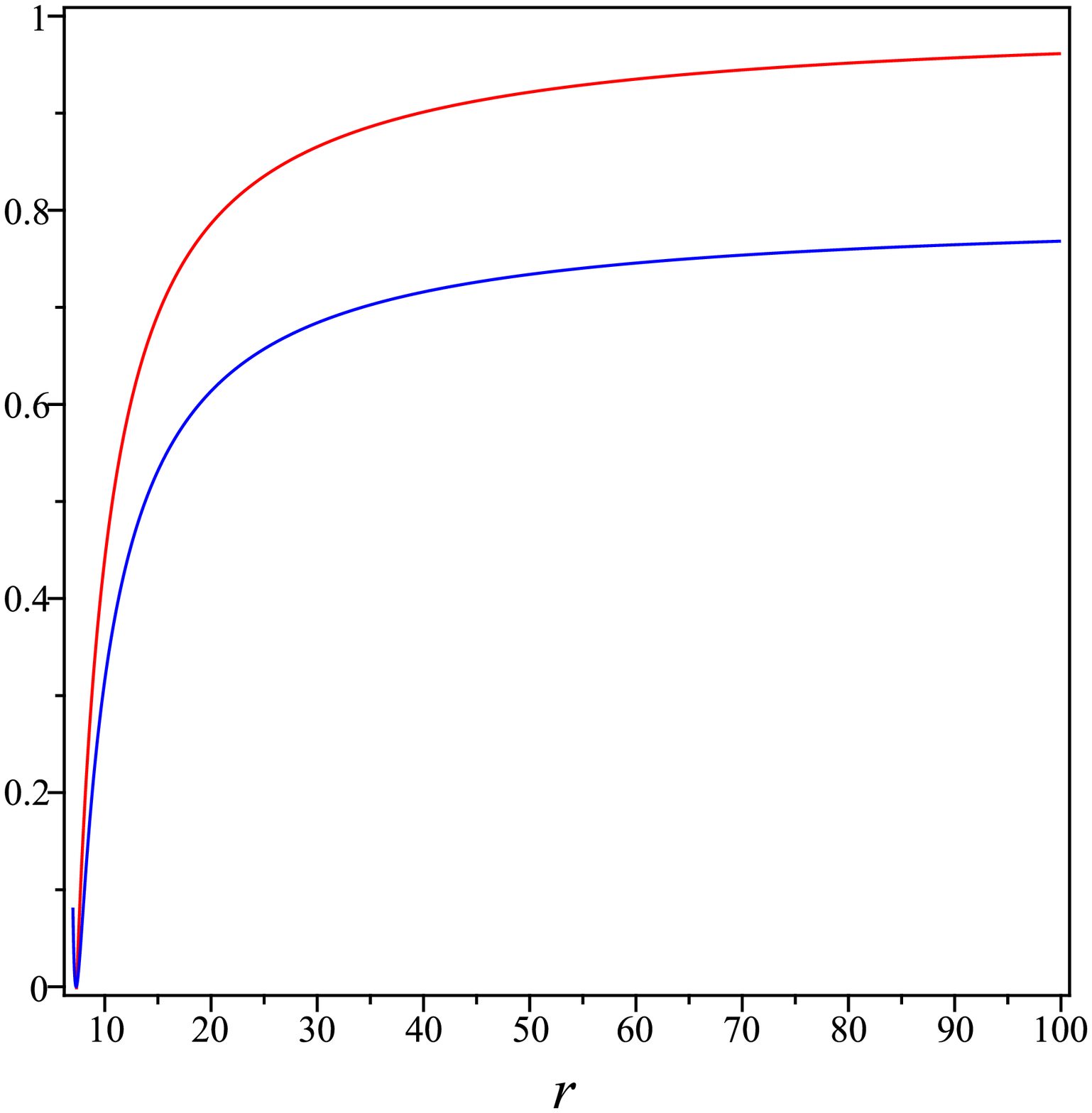}}
\subfigure[$h_{2}=2,
r_{+}=6$]{\includegraphics[width=0.3\columnwidth]{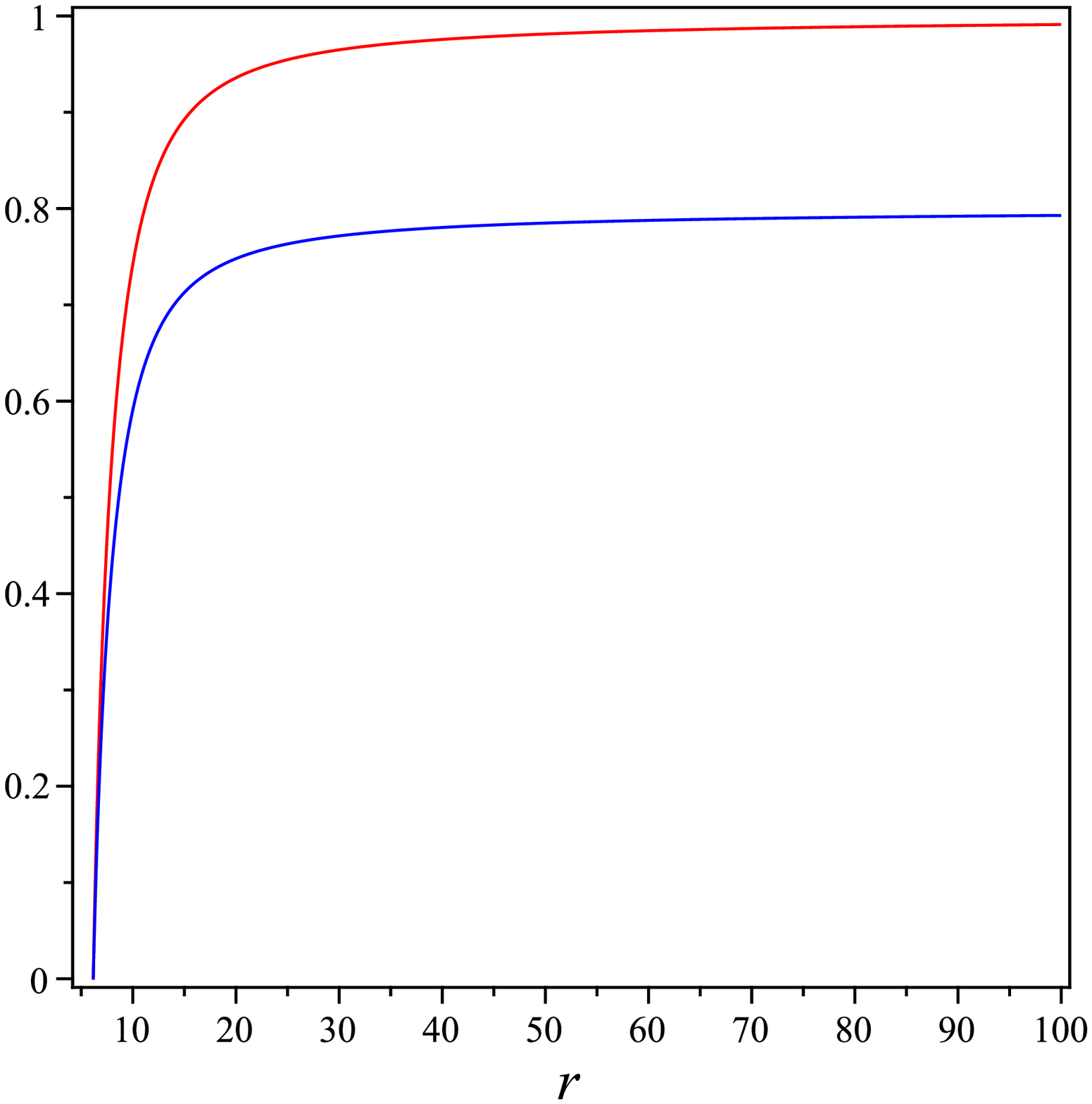}}
\subfigure[$h_{2}=1,r_{+}=2$]{\includegraphics[width=0.3\columnwidth]{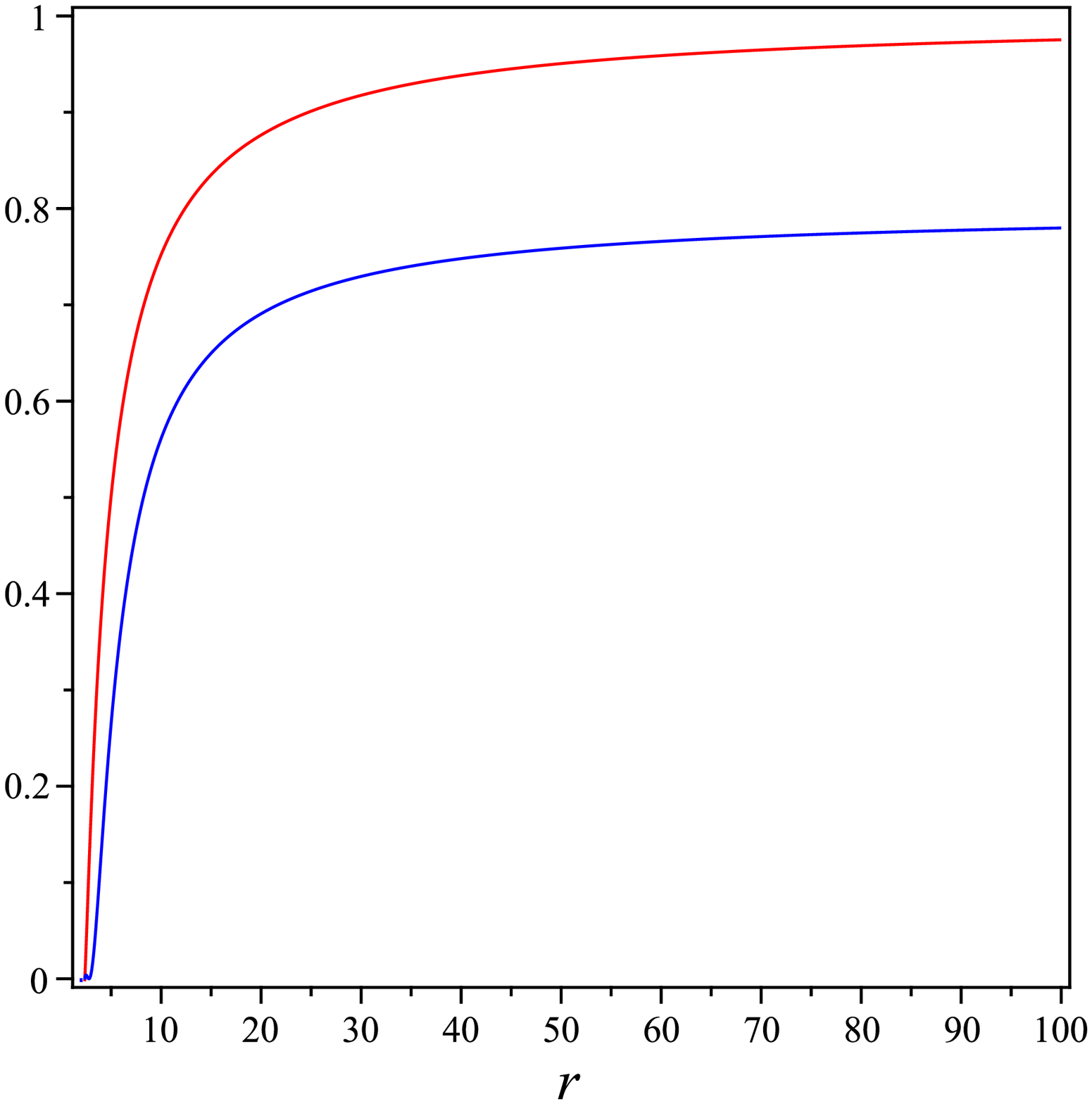}}
\caption{The behavior of $f(r)$ (blue line) and $0.8h(r)$ (red
line) in terms of $r$  for $c_{1}=\alpha=-0.5$. Left: correspond
to the red branch. Middle: correspond to the blue branch. Right:
correspond to the yellow branch in figures \ref{TMSplott}.}
\label{FHplott}
\end{figure}

In panels of Fig. \ref{FHplott}, we have depicted the metric
functions. In these figures, the mass of the black hole is
positive and these solutions are similar to Schwarzschild's black
hole.

\section{Conclusion}\label{con}

In this paper, we studied the black hole solutions of Einsteinian
cubic gravities by using continued fraction approximations. To get
a complete solution, first, we constructed the near horizon and
then asymptotic solutions and then used them to obtain an
approximate analytic solution using a continued-fraction
expansion. Then, we calculated the thermodynamic quantities like
entropy, temperature and mass, and by inserting them in the first
law and Smarr formula, we obtained the analytic solutions for the
near horizon quantities. Then, one can obtain
a metric according to continued fraction expansion that is only a function of constant integration not extra function like $f_{1}$ or $h_{1}$. We also showed that
continued fraction expansion can be used to accurately approximate
black hole solutions in cubic gravity, which are valid everywhere
outside of the event horizon. Finally, to obtain the new black
hole solutions, we considered the different relationships between
the near horizon constants. We also compared our results with
those of previous works on the subject and we found a good
agreement between them.\\
 The important point in our work is that
we assumed the near horizon constant $f_{1}$ in the first law of thermodynamic and the Smarr formula, is a function of the event horizon radius. This assumption is correct, because $f_{1}$ is proportional to temperature according to Equation \eqref{eq20}. This method is different from the one used in previous papers like \cite{Bueno:2016lrh}-\cite{Hennigar:2018hza}. First, they used the differential equations for metric function which have been obtained from the action of the theory, while, we have used the components of the field equations (\ref{fieldeq3}) of the theory. Second, they obtained the thermodynamical quantities from the first and the second term of near horizon expansion, while we obtained from the first law of thermodynamics and Smarr formula. Our method is applicable to every theory of gravity (as we previously applied to the quadratic gravity in reference \cite{Sajadi:2020axg}), while the method of the papers is only applicable for the theory in which the on-shell action is integrable with respect to $r$. Finally, this approach also allows us to obtain black hole solutions different from Einstein gravity solutions by assuming $f_{1}\neq h_{1}$ in the near horizon expansion as we have done in section (\ref{sec3}). We think using the thermodynamics to obtain the black hole solutions is a normal approach with respect to one in the references \cite{Bueno:2016lrh}-\cite{Hennigar:2018hza}.

We leave for future work, obtaining the non-vacuum, rotating black
hole, and other solutions of the theory by using the continued
fraction expansion.

\section*{Acknowledgements}
We thank the support of Iran National Science Foundation 99022223.

\appendix

\section{Explicit Terms in the Continued Fraction Approximation}\label{appa}

We present terms up to fourth order in the continued fraction
approximation:
\begin{align}
&\epsilon=-\dfrac{F_{1}}{r_{+}}-1,\,\,\,\, a_{1}=-1-a_{0}+2\epsilon+r_{+}f_{1},\,\,\,\, a_{2}=-{\dfrac {4a_{1}-5\epsilon+1+3 a_{0}+ f_{2}r_{+}^{2}}{{ a_{1}}}}
\nonumber \\
&a_{3}=-\dfrac{1}{{a_{1}}{a_{2}}}[-{f_{3}}{{r_{+}}}^{3}+{a_{1}}{{a_{2}}}^{2}+5{a_{1}}{a_{2}}+6{a_{0}}+10{a_{1}}-9\epsilon+1]
\label{cfrac-a}\\
&a_{4}=-\dfrac{{f_{4}}{{r_{+}}}^{4}+{a_{1}}{{a_{2}}}^{3}+2{a_{1}}{{a_{2}}}^{2}{a_{3}}+{a_{1}}{a_{2}}{{a_{3}}}^{2}+6{a_{1}}{{a_{2}}}^{2}+6{a_{1}}{a_{2}}{a_{3}}+15{a_{1}}{a_{2}}+10{a_{0}}+20{a_{1}} -14\epsilon+1}{{a_{1}}{a_{2}}{a_{3}}} \nonumber
\end{align}

\section{Near Horizon Constants}\label{appb}
Here, we present some near horizon constants regarding section (\ref{sec2}) and (\ref{sec3}) as follows:\\
The quantity  $f_4$ is given in \eqref{eq9}
\begin{align}
f_{4}&=\dfrac{-1}{3[{{r_{+}}^{10}
 ({r_{+}}^{3}+48\,\alpha\,f_{1}+24\,{r_{+}}\,\alpha\,{f_{{
1}}}^{2}) (48\,{r_{+}}\,\alpha\,{f_{1}}^{2}+96\,
\alpha\,f_{1}+{r_{+}}^{3}) }]}[-31\,{r_{+}}^{12}+7524\,{r_{+}}^{10}{f_{1}}^{2}
\alpha\nonumber\\
&-11064\,{ r_{+}}^{11}{f_{1}}^{3}\alpha- 4608\,{r_{+}}^{6}{
\alpha}^{2}{f_{1}}^{2}-86400\,{r_{+}}^{4}{\alpha}^{3}{f_{1}}^{4
}+10944\,{r_{+}}^{7}{f_{1}}^{3}{\alpha}^{2}-1534464\,{r_{+}}^{
2}{\alpha}^{4}{f_{1}}^{6}+\nonumber\\
&888192\,{r_{+}}^{5}{\alpha}^{3}{f_{1}
}^{5}-89136\,{r_{+}}^{8}{f_{1}}^{4}{\alpha}^{2}-5971968\,{\alpha}
^{5}{f_{1}}^{8}+5640192\,{\alpha}^{4}{f_{1}}^{7}{r_{+}}^{3}-
1769472\,{\alpha}^{3}{f_{1}}^{6}{r_{+}}^{6}+\nonumber\\
&184896\,{r_{+}}^{9
}{f_{1}}^{5}{\alpha}^{2}+1248\,\alpha\,{r_{+}}^{9}f_{1}+34\,{
r_{+}}^{13}f_{1}-864\,\alpha\,{r_{+}}^{8}].
\end{align}
The near horizon constants regarding section (\ref{sec3}):
\begin{align}
&f_{3}=-\frac {1}{2592({{f_{{1
}}}^{3}{h_{{1}}}^{2}{{\it r_{+}}}^{2}{\alpha}^{2} \left( {{\it r_{+}}}^{3}{f
_{{1}}}^{3}+8+6\,{{\it r_{+}}}^{2}{f_{{1}}}^{2}+12\,{\it r_{+}}\,f_{{1}}
 \right) })}[6912{\alpha}^{2}{f_{{1}}}^{6}{h_{{2}}}^{
2}{{\it r_{+}}}^{4}+71\,{h_{{1}}}^{2}{{\it r_{+}}}^{7}{f_{{1}}}^{3}-\nonumber\\
&160\,{{
\it r_{+}}}^{6}{h_{{1}}}^{2}{f_{{1}}}^{2}+59\,{{\it r_{+}}}^{5}{h_{{1}}}^{2}
f_{{1}}+30\,{{\it r_{+}}}^{4}{h_{{1}}}^{2}+5472\,{\alpha}^{2}{f_{{1}}}^{6
}{h_{{1}}}^{2}{{\it r_{+}}}^{2}+4536\,{\alpha}^{2}{f_{{1}}}^{4}{h_{{1}}}^
{2}+8892\,{\alpha}^{2}{f_{{1}}}^{5}{h_{{1}}}^{2}{\it r_{+}}+\nonumber\\
&876\,{f_{{1}}
}^{4}{h_{{1}}}^{2}{{\it r_{+}}}^{4}\alpha+816\,{f_{{1}}}^{5}{h_{{1}}}^{2}
{{\it r_{+}}}^{5}\alpha-792\,{{\it r_{+}}}^{2}{h_{{1}}}^{2}\alpha\,{f_{{1}}}
^{2}+1728\,{\alpha}^{2}{f_{{1}}}^{7}{h_{{1}}}^{2}{{\it r_{+}}}^{3}+6912\,
{\alpha}^{2}{f_{{1}}}^{5}{h_{{2}}}^{2}{{\it r_{+}}}^{3}-\nonumber\\
&816\,h_{{1}}{{
\it r_{+}}}^{5}h_{{2}}{f_{{1}}}^{4}\alpha+1728\,{\alpha}^{2}{f_{{1}}}^{7}
{h_{{2}}}^{2}{{\it r_{+}}}^{5}+684\,{f_{{1}}}^{3}{h_{{1}}}^{2}{{\it r_{+}}}^
{3}\alpha+576\,h_{{1}}{{\it r_{+}}}^{3}h_{{2}}{f_{{1}}}^{2}\alpha-19584\,
{\alpha}^{2}{f_{{1}}}^{5}h_{{1}}{{\it r_{+}}}^{2}h_{{2}}+\nonumber\\
&2496\,h_{{1}}{{
\it r_{+}}}^{4}h_{{2}}{f_{{1}}}^{3}\alpha-960\,h_{{1}}{{\it r_{+}}}^{6}{f_{{
1}}}^{5}\alpha\,h_{{2}}-12384\,{\alpha}^{2}{f_{{1}}}^{6}h_{{1}}{{\it
r_{+}}}^{3}h_{{2}}-3456\,{\alpha}^{2}{f_{{1}}}^{7}h_{{1}}{{\it r_{+}}}^{4}h_
{{2}}-17280\,{\alpha}^{2}{f_{{1}}}^{4}h_{{2}}{\it r_{+}}\,h_{{1}}].
\end{align}

\begin{align}
&h_{3}=-\frac {1}{2592({{f_{{1}}}^{
4}h_{{1}}{{\it r_{+}}}^{2}{\alpha}^{2} \left( {{\it r_{+}}}^{3}{f_{{1}}}^{3}
+8+6\,{{\it r_{+}}}^{2}{f_{{1}}}^{2}+12\,{\it r_{+}}\,f_{{1}} \right) })}[11\,{h_{{1}}}^{2}{{\it r_{+}}}^{7}{f_{{1}}}^{
3}-40\,{{\it r_{+}}}^{6}{h_{{1}}}^{2}{f_{{1}}}^{2}+35\,{{\it r_{+}}}^{5}{h_{
{1}}}^{2}f_{{1}}\nonumber\\
&-6\,{{\it r_{+}}}^{4}{h_{{1}}}^{2}+7200\,{\alpha}^{2}{f_{
{1}}}^{6}{h_{{1}}}^{2}{{\it r_{+}}}^{2}+8424\,{\alpha}^{2}{f_{{1}}}^{4}{h
_{{1}}}^{2}+12348\,{\alpha}^{2}{f_{{1}}}^{5}{h_{{1}}}^{2}{\it r_{+}}+948
\,{f_{{1}}}^{4}{h_{{1}}}^{2}{{\it r_{+}}}^{4}\alpha+528\,{f_{{1}}}^{5}{h_
{{1}}}^{2}{{\it r_{+}}}^{5}\alpha\nonumber\\
&-360\,{{\it r_{+}}}^{2}{h_{{1}}}^{2}\alpha
\,{f_{{1}}}^{2}+1728\,{\alpha}^{2}{f_{{1}}}^{7}{h_{{1}}}^{2}{{\it r_{+}}}
^{3}+6912\,{\alpha}^{2}{f_{{1}}}^{5}{h_{{2}}}^{2}{{\it r_{+}}}^{3}+6912\,
{\alpha}^{2}{f_{{1}}}^{6}{h_{{2}}}^{2}{{\it r_{+}}}^{4}+1728\,{\alpha}^{2
}{f_{{1}}}^{7}{h_{{2}}}^{2}{{\it r_{+}}}^{5}+\nonumber\\
&252\,{f_{{1}}}^{3}{h_{{1}}}^
{2}{{\it r_{+}}}^{3}\alpha+576\,h_{{1}}{{\it r_{+}}}^{3}h_{{2}}{f_{{1}}}^{2}
\alpha-528\,h_{{1}}{{\it r_{+}}}^{5}h_{{2}}{f_{{1}}}^{4}\alpha+1920\,h_{{
1}}{{\it r_{+}}}^{4}h_{{2}}{f_{{1}}}^{3}\alpha-672\,h_{{1}}{{\it r_{+}}}^{6}
{f_{{1}}}^{5}\alpha\,h_{{2}}-\nonumber\\
&14112\,{\alpha}^{2}{f_{{1}}}^{6}h_{{1}}{{
\it r_{+}}}^{3}h_{{2}}-23040\,{\alpha}^{2}{f_{{1}}}^{5}h_{{1}}{{\it r_{+}}}^
{2}h_{{2}}-3456\,{\alpha}^{2}{f_{{1}}}^{7}h_{{1}}{{\it r_{+}}}^{4}h_{{2}}
-17280\,{\alpha}^{2}{f_{{1}}}^{4}h_{{2}}{\it r_{+}}\,h_{{1}}].
\end{align}

\begin{align}
&f_{4}=\frac {-1}{248832({{f_{{1}}}^{5}{h_{{1}}}^
{3}{{r_{+}}}^{3}{\alpha}^{3} \left( {{r_{+}}}^{5}{f_{{1}}}^{5}+10\,{
{r_{+}}}^{4}{f_{{1}}}^{4}+80\,{{r_{+}}}^{2}{f_{{1}}}^{2}+32+40\,{{
r_{+}}}^{3}{f_{{1}}}^{3}+80\,{r_{+}}\,f_{{1}} \right) })}\, [-922\,{h_{{1}}}^{3}{{ r_{+}}}^{7}f_{{1}}
\nonumber\\
&+2557\,{h_{{1}}}^{3}{r_{+}}^{11}{f_{1}}^{5}+7713\,{h_{1}}^{3}{
r_{+}}^{9}{f_{1}}^{3}-8528\,{h_{{1}}}^{3}{{r_{+}}}^{10}{f_{1}}^
{4}-328\,{h_{1}}^{3}{r_{+}}^{8}{f_{1}}^{2}+813888\,{\alpha}^{3}
{r_{+}}^{3}{f_{1}}^{8}h_{2}{h_{1}}^{2}-\nonumber\\
& 1990656\,{\alpha}^{3}{
{r_{+}}}^{5}{f_{1}}^{8}{h_{2}}^{3}+8136\,{f_{1}}^{5}{h_{1}}^{
3}{{r_{+}}}^{3}{\alpha}^{2}-59424\,{f_{1}}^{4}{h_{1}}^{3}{{r_{+}
}}^{6}\alpha -1492992\,{\alpha}^{3}{{ r_{+}}}^{7}{f_{1}}^{10}{h_{2}
}^{3}+\nonumber\\
& 247896\,{f_{{1}}}^{8}{h_{{1}}}^{3}{{r_{+}}}^{6}{\alpha}^{2}+
89712\,{f_{{1}}}^{6}{h_{{1}}}^{3}{{r_{+}}}^{4}{\alpha}^{2}-248832\,{
\alpha}^{3}{{r_{+}}}^{8}{f_{{1}}}^{11}{h_{{2}}}^{3}+143424\,{f_{{1}}}
^{9}{h_{{1}}}^{3}{{r_{+}}}^{7}{\alpha}^{2}+\nonumber\\
&384012\,{f_{{1}}}^{7}{h_{{
1}}}^{3}{{r_{+}}}^{5}{\alpha}^{2}+230688\,{\alpha}^{3}{f_{{1}}}^{8}{h
_{{1}}}^{3}{{r_{+}}}^{2}+248832\,{\alpha}^{3}{f_{{1}}}^{10}{h_{{1}}}^
{3}{{r_{+}}}^{4}+316224\,{\alpha}^{3}{f_{{1}}}^{9}{h_{{1}}}^{3}{{
r_{+}}}^{3}+\nonumber\\
&654480\,{\alpha}^{3}{f_{{1}}}^{7}{h_{{1}}}^{3}{r_{+}}+37164
\,{f_{{1}}}^{7}{h_{{1}}}^{3}{{r_{+}}}^{9}\alpha +31788\,{f_{{1}}}^{3}{
h_{{1}}}^{3}{{r_{+}}}^{5}\alpha -15924\,{f_{{1}}}^{6}{h_{{1}}}^{3}{{
r_{+}}}^{8}\alpha +4824\,{{r_{+}}}^{4}{h_{{1}}}^{3}\alpha\,{f_{{1}}}^
{2}\nonumber\\
&-6480\,{{r_{+}}}^{2}{h_{{1}}}^{3}{\alpha}^{2}{f_{{1}}}^{4}-7608\,{
h_{{1}}}^{3}{{r_{+}}}^{7}{f_{{1}}}^{5}\alpha -2985984\,{\alpha}^{3}{{
r_{+}}}^{6}{f_{{1}}}^{9}{h_{{2}}}^{3}+124416\,{\alpha}^{3}{f_{{1}}}^{
11}{h_{{1}}}^{3}{{r_{+}}}^{5}-\nonumber\\
&288000\,{\alpha}^{2}{{r_{+}}}^{6}{f_{{
1}}}^{7}h_{{2}}{h_{{1}}}^{2}-492\,{{r_{+}}}^{6}{h_{{1}}}^{3}+676512\,
{\alpha}^{3}{f_{{1}}}^{6}{h_{{1}}}^{3}+237312\,{\alpha}^{2}{{r_{+}}}^
{9}{f_{{1}}}^{9}{h_{{2}}}^{2}h_{{1}}-16128\,\alpha\,{f_{{1}}}^{2}h_{{2
}}{{r_{+}}}^{5}{h_{{1}}}^{2}\nonumber\\
&-829440\,{\alpha}^{3}{f_{{1}}}^{6}{h_{{2}
}}^{2}{{r_{+}}}^{2}h_{{1}}-138240\,{\alpha}^{2}{f_{{1}}}^{4}{h_{{2}}}
^{2}{{r_{+}}}^{4}h_{{1}}-399168\,{\alpha}^{3}{{r_{+}}}^{4}{f_{{1}}}^
{9}h_{{2}}{h_{{1}}}^{2}-87024\,\alpha\,{{r_{+}}}^{7}{f_{{1}}}^{4}h_{{
2}}{h_{{1}}}^{2}+\nonumber\\
&1302912\,{\alpha}^{3}{{r_{+}}}^{2}{f_{{1}}}^{7}h_{{2
}}{h_{{1}}}^{2}-356544\,{\alpha}^{2}{{r_{+}}}^{8}{f_{{1}}}^{9}h_{{2}}
{h_{{1}}}^{2}+580608\,{\alpha}^{3}{{r_{+}}}^{7}{f_{{1}}}^{11}{h_{{2}}
}^{2}h_{{1}}+2377728\,{\alpha}^{3}{{r_{+}}}^{3}{f_{{1}}}^{7}{h_{{2}}}
^{2}h_{{1}}\nonumber\\
&-414720\,{\alpha}^{3}{{r_{+}}}^{6}{f_{{1}}}^{11}h_{{2}}{h_
{{1}}}^{2}+2598912\,{\alpha}^{3}{{\it r_{+}}}^{6}{f_{{1}}}^{10}{h_{{2}}}^
{2}h_{{1}}+4230144\,{\alpha}^{3}{{\it r_{+}}}^{5}{f_{{1}}}^{9}{h_{{2}}}^{
2}h_{{1}}-460800\,{\alpha}^{2}{{\it r_{+}}}^{5}{f_{{1}}}^{5}{h_{{2}}}^{2}
h_{{1}}\nonumber\\
&-34560\,{\alpha}^{2}{{\it r_{+}}}^{7}{f_{{1}}}^{7}{h_{{2}}}^{2}h_{
{1}}-1260288\,{\alpha}^{2}{{\it r_{+}}}^{6}{f_{{1}}}^{6}{h_{{2}}}^{2}h_{{
1}}+4105728\,{\alpha}^{3}{{\it r_{+}}}^{4}{f_{{1}}}^{8}{h_{{2}}}^{2}h_{{1
}}+189360\,\alpha\,{{\it r_{+}}}^{8}{f_{{1}}}^{5}h_{{2}}{h_{{1}}}^{2}+\nonumber\\
&
9072\,\alpha\,{{\it r_{+}}}^{9}{f_{{1}}}^{6}h_{{2}}{h_{{1}}}^{2}-52128\,
\alpha\,{{\it r_{+}}}^{6}{f_{{1}}}^{3}h_{{2}}{h_{{1}}}^{2}+110592\,{
\alpha}^{2}{f_{{1}}}^{4}h_{{2}}{{\it r_{+}}}^{3}{h_{{1}}}^{2}-746496\,{
\alpha}^{3}{f_{{1}}}^{6}h_{{2}}{\it r_{+}}\,{h_{{1}}}^{2}-\nonumber\\
&771264\,{\alpha
}^{2}{{\it r_{+}}}^{7}{f_{{1}}}^{8}h_{{2}}{h_{{1}}}^{2}-50928\,\alpha\,{{
\it r_{+}}}^{10}{f_{{1}}}^{7}h_{{2}}{h_{{1}}}^{2}+313920\,{\alpha}^{2}{{
\it r_{+}}}^{5}{f_{{1}}}^{6}h_{{2}}{h_{{1}}}^{2}-1022976\,{\alpha}^{3}{{
\it r_{+}}}^{5}{f_{{1}}}^{10}h_{{2}}{h_{{1}}}^{2}+\nonumber\\
&1369728\,{\alpha}^{2}{{
\it r_{+}}}^{4}{f_{{1}}}^{5}h_{{2}}{h_{{1}}}^{2}+723456\,{\alpha}^{2}{{
\it r_{+}}}^{8}{f_{{1}}}^{8}{h_{{2}}}^{2}h_{{1}}].
\end{align}

\begin{align}
&h_{4}=\frac {-1}{248832({{f_{{1}}}^{6}{h_{{1}}}^
{2}{{\it r_{+}}}^{3}{\alpha}^{3} \left( {{\it r_{+}}}^{5}{f_{{1}}}^{5}+10\,{
{\it r_{+}}}^{4}{f_{{1}}}^{4}+80\,{{\it r_{+}}}^{2}{f_{{1}}}^{2}+32+40\,{{
\it r_{+}}}^{3}{f_{{1}}}^{3}+80\,{\it r_{+}}\,f_{{1}} \right) })}\, [-810\,{h_{{1}}}^{3}{{\it r_{+}}}^{7}f_{{1}}
\nonumber\\
&+709\,{h_{{1}}}^{3}{{\it r_{+}}}^{11}{f_{{1}}}^{5}+3177\,{h_{{1}}}^{3}{{
\it r_{+}}}^{9}{f_{{1}}}^{3}-2832\,{h_{{1}}}^{3}{{\it r_{+}}}^{10}{f_{{1}}}^
{4}-424\,{h_{{1}}}^{3}{{\it r_{+}}}^{8}{f_{{1}}}^{2}+426816\,{\alpha}^{3}
{{\it r_{+}}}^{3}{f_{{1}}}^{8}h_{{2}}{h_{{1}}}^{2}-\nonumber\\
&1990656\,{\alpha}^{3}{
{\it r_{+}}}^{5}{f_{{1}}}^{8}{h_{{2}}}^{3}-251640\,{f_{{1}}}^{5}{h_{{1}}}
^{3}{{\it r_{+}}}^{3}{\alpha}^{2}-36576\,{f_{{1}}}^{4}{h_{{1}}}^{3}{{\it
r_{+}}}^{6}\alpha -1492992\,{\alpha}^{3}{{\it r_{+}}}^{7}{f_{{1}}}^{10}{h_{{2
}}}^{3}+\nonumber\\
&354456\,{f_{{1}}}^{8}{h_{{1}}}^{3}{{\it r_{+}}}^{6}{\alpha}^{2}+
126576\,{f_{{1}}}^{6}{h_{{1}}}^{3}{{\it r_{+}}}^{4}{\alpha}^{2}-248832\,{
\alpha}^{3}{{\it r_{+}}}^{8}{f_{{1}}}^{11}{h_{{2}}}^{3}+129600\,{f_{{1}}}
^{9}{h_{{1}}}^{3}{{\it r_{+}}}^{7}{\alpha}^{2}+\nonumber\\
&442476\,{f_{{1}}}^{7}{h_{{
1}}}^{3}{{\it r_{+}}}^{5}{\alpha}^{2}-377568\,{\alpha}^{3}{f_{{1}}}^{8}{h
_{{1}}}^{3}{{\it r_{+}}}^{2}+331776\,{\alpha}^{3}{f_{{1}}}^{10}{h_{{1}}}^
{3}{{\it r_{+}}}^{4}+205632\,{\alpha}^{3}{f_{{1}}}^{9}{h_{{1}}}^{3}{{\it
r_{+}}}^{3}-\nonumber\\
&482544\,{\alpha}^{3}{f_{{1}}}^{7}{h_{{1}}}^{3}{\it r_{+}}+24780
\,{f_{{1}}}^{7}{h_{{1}}}^{3}{{\it r_{+}}}^{9}\alpha+11532\,{f_{{1}}}^{3}{
h_{{1}}}^{3}{{\it r_{+}}}^{5}\alpha+11628\,{f_{{1}}}^{6}{h_{{1}}}^{3}{{
\it r_{+}}}^{8}\alpha+6552\,{{\it r_{+}}}^{4}{h_{{1}}}^{3}\alpha\,{f_{{1}}}^
{2}\nonumber\\
&-113616\,{{\it r_{+}}}^{2}{h_{{1}}}^{3}{\alpha}^{2}{f_{{1}}}^{4}-26232
\,{h_{{1}}}^{3}{{\it r_{+}}}^{7}{f_{{1}}}^{5}\alpha-2985984\,{\alpha}^{3}
{{\it r_{+}}}^{6}{f_{{1}}}^{9}{h_{{2}}}^{3}+124416\,{\alpha}^{3}{f_{{1}}}
^{11}{h_{{1}}}^{3}{{\it r_{+}}}^{5}-\nonumber\\
&472320\,{\alpha}^{2}{{\it r_{+}}}^{6}{f_
{{1}}}^{7}h_{{2}}{h_{{1}}}^{2}+180\,{{\it r_{+}}}^{6}{h_{{1}}}^{3}+116640
\,{\alpha}^{3}{f_{{1}}}^{6}{h_{{1}}}^{3}+195840\,{\alpha}^{2}{{\it r_{+}}
}^{9}{f_{{1}}}^{9}{h_{{2}}}^{2}h_{{1}}-11520\,\alpha\,{f_{{1}}}^{2}h_{
{2}}{{\it r_{+}}}^{5}{h_{{1}}}^{2}\nonumber\\
&-829440\,{\alpha}^{3}{f_{{1}}}^{6}{h_{{
2}}}^{2}{{\it r_{+}}}^{2}h_{{1}}-138240\,{\alpha}^{2}{f_{{1}}}^{4}{h_{{2}
}}^{2}{{\it r_{+}}}^{4}h_{{1}}-1200960\,{\alpha}^{3}{{\it r_{+}}}^{4}{f_{{1}
}}^{9}h_{{2}}{h_{{1}}}^{2}-54768\,\alpha\,{{\it r_{+}}}^{7}{f_{{1}}}^{4}h
_{{2}}{h_{{1}}}^{2}+\nonumber\\
&1496448\,{\alpha}^{3}{{\it r_{+}}}^{2}{f_{{1}}}^{7}h_
{{2}}{h_{{1}}}^{2}-301248\,{\alpha}^{2}{{\it r_{+}}}^{8}{f_{{1}}}^{9}h_{{
2}}{h_{{1}}}^{2}+580608\,{\alpha}^{3}{{\it r_{+}}}^{7}{f_{{1}}}^{11}{h_{{
2}}}^{2}h_{{1}}+2377728\,{\alpha}^{3}{{\it r_{+}}}^{3}{f_{{1}}}^{7}{h_{{2
}}}^{2}h_{{1}}-\nonumber\\
&414720\,{\alpha}^{3}{{\it r_{+}}}^{6}{f_{{1}}}^{11}h_{{2}}
{h_{{1}}}^{2}+2847744\,{\alpha}^{3}{{\it r_{+}}}^{6}{f_{{1}}}^{10}{h_{{2}
}}^{2}h_{{1}}+5225472\,{\alpha}^{3}{{\it r_{+}}}^{5}{f_{{1}}}^{9}{h_{{2}}
}^{2}h_{{1}}-460800\,{\alpha}^{2}{{\it r_{+}}}^{5}{f_{{1}}}^{5}{h_{{2}}}^
{2}h_{{1}}-\nonumber\\
&34560\,{\alpha}^{2}{{\it r_{+}}}^{7}{f_{{1}}}^{7}{h_{{2}}}^{2}
h_{{1}}-1094400\,{\alpha}^{2}{{\it r_{+}}}^{6}{f_{{1}}}^{6}{h_{{2}}}^{2}h
_{{1}}+5101056\,{\alpha}^{3}{{\it r_{+}}}^{4}{f_{{1}}}^{8}{h_{{2}}}^{2}h_
{{1}}+124848\,\alpha\,{{\it r_{+}}}^{8}{f_{{1}}}^{5}h_{{2}}{h_{{1}}}^{2}+
\nonumber\\
&8688\,\alpha\,{{\it r_{+}}}^{9}{f_{{1}}}^{6}h_{{2}}{h_{{1}}}^{2}-43680\,
\alpha\,{{\it r_{+}}}^{6}{f_{{1}}}^{3}h_{{2}}{h_{{1}}}^{2}+276480\,{
\alpha}^{2}{f_{{1}}}^{4}h_{{2}}{{\it r_{+}}}^{3}{h_{{1}}}^{2}+82944\,{
\alpha}^{3}{f_{{1}}}^{6}h_{{2}}{\it r_{+}}\,{h_{{1}}}^{2}-\nonumber\\
&810432\,{\alpha
}^{2}{{\it r_{+}}}^{7}{f_{{1}}}^{8}h_{{2}}{h_{{1}}}^{2}-31344\,\alpha\,{{
\it r_{+}}}^{10}{f_{{1}}}^{7}h_{{2}}{h_{{1}}}^{2}+489024\,{\alpha}^{2}{{
\it r_{+}}}^{5}{f_{{1}}}^{6}h_{{2}}{h_{{1}}}^{2}-1354752\,{\alpha}^{3}{{
\it r_{+}}}^{5}{f_{{1}}}^{10}h_{{2}}{h_{{1}}}^{2}+\nonumber\\
&1342080\,{\alpha}^{2}{{
\it r_{+}}}^{4}{f_{{1}}}^{5}h_{{2}}{h_{{1}}}^{2}+599040\,{\alpha}^{2}{{
\it r_{+}}}^{8}{f_{{1}}}^{8}{h_{{2}}}^{2}h_{{1}}].
\end{align}
{
\section{Functions}\label{app3}
Here, we present the functions regarding differential equations (\ref{eqasymp45}) and (\ref{eqasymp46}) as follows:\\
\begin{align}
\gamma(r)&=\dfrac{4r^{7}-9Mr^{6}-528\alpha M^{2}r+816\alpha M^{3}}{r(r-2M)(r^{6}+240\alpha M^{2})},\\
\eta(r)&=\dfrac{2r^{7}-3Mr^{6}-528\alpha M^{2}r+1296\alpha M^{3}}{r(r-2M)(r^{6}+240\alpha M^{2})},\\
\omega(r)&=\dfrac{2(2r^{8}-6Mr^{7}+3M^{2}r^{6}+240\alpha M^{2}r^{2}-1968\alpha M^{3}r+2736\alpha M^{4})}{r^{2}(2M-r)^{2}(r^{6}+240\alpha M^{2})},\\
\Xi(r)&=-\dfrac{2M(2r^{7}-5Mr^{6}-1008\alpha M^{2}r+1776\alpha M^{3})}{r^{2}(r^{6}+240\alpha M^{2})(2M-r)^{2}},\\
g(r)&=-\dfrac{1344\alpha M^{3}}{r^{3}(r^{6}+240\alpha M^{2})}.
\end{align}
and
\begin{align}
\bar{\gamma}(r)&=\dfrac{2r^{7}-3Mr^{6}-352\alpha M^{2}r+864\alpha M^{3}}{r(r-2M)(r^{6}+160\alpha M^{2})},\\
\bar{\eta}(r)&=\dfrac{2r^{7}-5Mr^{6}-352\alpha M^{2}r+544\alpha M^{3}}{r(r-2M)(r^{6}+160\alpha M^{2})},\\
\bar{\omega}(r)&=\dfrac{2M^{2}(r^{6}+672r\alpha M-1184\alpha M^{2})}{r^{2}(r^{6}+160\alpha M^{2})(2M-r)^{2}},\\
\bar{\Xi}(r)&=\dfrac{2(r^{8}-4Mr^{7}+3M^{2}r^{6}+160\alpha M^{2}r^{2}-1312\alpha M^{3}r+1824\alpha M^{4})}{r^{2}(r^{6}+160\alpha M^{2})(2M-r)^{2}},\\
\bar{g}(r)&=-\dfrac{896\alpha M^{3}}{r^{3}(r^{6}+160\alpha M^{2})}.
\end{align}}

\end{document}